\documentclass[floatfix,secnumarabic,amssymb,nobibnotes,nofootinbib,preprint,aps,pra, superscriptaddress,showpacs]{revtex4-2}

\usepackage{amsmath}
\usepackage{amsfonts}
\usepackage{amssymb}
\usepackage{graphicx}
\usepackage{units}
\usepackage{color}
\usepackage{xspace}
\usepackage[normalem]{ulem}
\usepackage{natbib}
\usepackage{textcomp}

\newcommand{\fig}[1]{Fig.~\ref{#1}}

\DeclareUnicodeCharacter{2009}{\,}
\DeclareUnicodeCharacter{2212}{-}

\makeatletter
\def\@fnsymbol#1{\ensuremath{\ifcase#1\or \dagger\or \star\or
   \mathsection\or \mathparagraph\or \|\or **\or \star
   \or \star \else\@ctrerr\fi}}
    \makeatother

\begin{document}
\title{In-Plane Anisotropy-Driven Directional Charge Transport in van der Waals \textit{p-n} Heterojunction}

\author{Rahul Paramanik}
\affiliation{School of Physical Sciences, Indian Association for the Cultivation of Science, 2A $\&$ B 
Raja S. C. Mullick Road, Jadavpur, Kolkata- 700032, India}

\author{Tanima Kundu}
\affiliation{School of Physical Sciences, Indian Association for the Cultivation of Science, 2A $\&$ B 
Raja S. C. Mullick Road, Jadavpur, Kolkata- 700032, India}

\author{Soumik Das}
\affiliation{School of Physical Sciences, Indian Association for the Cultivation of Science, 2A $\&$ B 
Raja S. C. Mullick Road, Jadavpur, Kolkata- 700032, India}

\author{Alexey Barinov}
\affiliation{Sincrotrone Trieste s.c.p.a., 34149 Basovizza, Trieste, Italy}

\author{Bikash Das}
\affiliation{School of Physical Sciences, Indian Association for the Cultivation of Science, 2A $\&$ B 
Raja S. C. Mullick Road, Jadavpur, Kolkata- 700032, India}

\author{Bipul Karmakar}
\affiliation{School of Physical Sciences, Indian Association for the Cultivation of Science, 2A $\&$ B 
Raja S. C. Mullick Road, Jadavpur, Kolkata- 700032, India}

\author{Sujan Maity}
\affiliation{School of Physical Sciences, Indian Association for the Cultivation of Science, 2A $\&$ B 
Raja S. C. Mullick Road, Jadavpur, Kolkata- 700032, India}

\author{Mainak Palit}
\affiliation{School of Physical Sciences, Indian Association for the Cultivation of Science, 2A $\&$ B 
Raja S. C. Mullick Road, Jadavpur, Kolkata- 700032, India}

\author{Kapildeb Dolui}
\affiliation{Department of Physics, Indian Institute of Technology Tirupati, Tirupati, Andhra Pradesh 517619, India}

\author{Sanjoy Kr Mahatha}
\email{sanjoymahatha@gmail.com} 
\affiliation{UGC-DAE Consortium for Scientific Research, Khandwa Road, Indore 452001, Madhya Pradesh, India}

\author{Subhadeep Datta}
\email{sspsdd@iacs.res.in}
\affiliation{School of Physical Sciences, Indian Association for the Cultivation of Science, 2A $\&$ B 
Raja S. C. Mullick Road, Jadavpur, Kolkata- 700032, India}

\begin{abstract}

\textbf{
Low-symmetry two-dimensional (2D) van der Waals (vdW) materials enable anisotropic charge transport, crucial for polarization-sensitive optoelectronics. In this study, a $p$-GeS/$n$-MoS$_2$ heterostructure diode is investigated, where the anisotropic band dispersion of GeS, revealed by angle-resolved photoemission spectroscopy (ARPES), governs directional charge flow. Angle-resolved Raman spectroscopy confirms the crystallographic orientation, and transport measurements in GeS field-effect transistors (FETs) show a mobility anisotropy of $\sim 3.4$. The heterojunction exhibits orientation-dependent diode characteristics, anti-ambipolar transport, and a type-II band alignment, leading to anisotropic optoelectronic response. These findings establish a pathway for utilizing electronic anisotropy in vdW heterostructures for energy-efficient rectification.} 

\end{abstract}

\maketitle

\section*{Introduction}

Directional charge transport in low-symmetry materials originates from intrinsic anisotropic mobility (\(\mu_x \neq \mu_y\), with \(\mu_x\) and \(\mu_y\) representing carrier mobility along principal in-plane axes), resulting in orientation-dependent current flow \cite{tomanek}. In the heterostructure framework, rectification under polarized excitation enables intrinsic diode behavior for high-frequency applications \cite{zhang}. Recently, Low-symmetry layered materials with their intrinsic in-plane anisotropy have attracted significant attention for excellent multifunctional capabilities in next-generation electronic and optoelectronic device physics \cite{review-1,review-2,review-3}. The family of in-plane anisotropic 2D materials constitutes black phosphorous \cite{BP}, distorted phase of transition metal dichalcogenides like 1T$^{’}$ WTe$_2$ \cite{WTe2-1}, ReS$_2$ \cite{ReS2-1}, ReSe$_2$ \cite{ReSe2-1}, transition metal trichalcogenides e.g. TiS$_3$ \cite{TiS3-1}, puckered structure of PdSe$_2$ \cite{PdSe2-1}, group IV-V compounds such as GeP \cite{GeP}, GeAs \cite{GeAs}, newly discovered penta-PdPSe \cite{PdPSe} etc. Compared to other materials, this family especially presents unique optical, electrical and thermal properties such as, tunable band gap, anisotropic in-plane optical conductivity, anisotropic light absorption and high mobility, which open up comprehensive performances as polarization sensitive photodetectors \cite{app-1,app-2}, linearly polarized pulse generators \cite{app-3}, polarization sensors \cite{app-4}, digital inverters \cite{app-5}, artificial synaptic devices \cite{app-6}, and so on. In spite of so many excellent features, most of them suffer from air-instability and low ON/OFF ratio restricting their applications in future electronics \cite{BP-air}. Combining a group IV element (Ge or Sn) with group V-VI elements (P, As, S and Se) have been revived with recent advancement incorporating high air-stability, strong in-plane anisotropy, high mobility and excellent device performance \cite{GeP,GeAs,GeS,GeSe}. 
In particular materials such as SnSe \cite{ACSAMI_2019}, SnS \cite{Nanoscale_2022}, SnS$_{1-x}$Se$_x$ \cite{ACSNano_2020_SnSSe,Nanoscale_2021_SnSSe,ACS_AMI_2024}, and GeS$_{1-x}$Se$_x$ \cite{AdvMater_2024_GeSSe} have demonstrated their remarkable potential in electronic and optoelectronic devices, further establishing this material family as highly desirable for solar cells and photodetectors.

Germanium sulfide (GeS) was explored as a 2D monochalcogenide in the family of low-symmetry anisotropic materials \cite{GeS}. Bulk GeS exhibits $p$- type semiconducting behavior with 1.6 eV indirect band gap. The unit cell of GeS covers two adjacent layers with threefold coordination of each germanium. The anisotropic nature is attributed by its puckered orthorhombic structure with armchair (AC) and zigzag (ZZ) atomic arrangements of two different in-plane lattice directions. Angle-dependent photoluminescence spectra and absorption spectroscopy reveal the anisotropic layered structure of GeS \cite{GeS,GeS-opt}. In recent years, GeS is recognized as a promising candidate in designing high-efficiency solar cells and photodetectors \cite{GeS-app,nanoscale-2016}.

To utilize the anisotropic geometry coupled with photo-tunable characteristics, the most effective strategy is to construct vdW heterostructure for opto-electrical modulation of the mutual band alignment. In contrast to the conventional covalently or ionic bonded heterostructure, vdW heterostructure (vdWH) can be easily formed without any constraint of lattice matching and provides an ideal platform for multifunctional device applications \cite{NatMat_2013,Science_2012,NatNano_2013,APL_2013,Nature_2019}. So far, vdWHs aiming at rectifying $p$-$n$ junction diodes and photovoltaic detectors have been extensively studied in the last few years \cite{ACSNano_2014,Nanoscale_2015,Nanolett_2015,advmat_2016,GeSe_MoS2_pn,nanoscale-2023}. Also, the photoconducting capability of the anisotropic layered materials are utilized in constructing vdWH-based photodiodes \cite{AdFM_2019,ACS-AMI_2020,ACS-AMI_2024}. But the effect of anisotropy is not well-explored in terms of diode characteristics. 

Herein, we present a comprehensive study of angle-resolved photoemission spectroscopy (ARPES) and field-effect transport (FET) in GeS to elucidate the role of in-plane anisotropy in its electronic structure and device performance. ARPES spectra of freshly cleaved GeS single crystals reveal an anisotropic band dispersion along orthogonal crystallographic directions, in agreement with prior reports \cite{arpes-1,arpes-2} and density functional theory (DFT) calculations. FET measurements further confirm directional-dependent electrical transport, with the highest drain current along the armchair direction, yielding a maximum hole mobility of 0.038 cm$^{2}$/V·s and an ON/OFF ratio of $\sim 10^3$. Angle-resolved polarized Raman spectroscopy (ARPRS) is employed to determine crystal orientations, crucial for leveraging anisotropy in optoelectronic applications. To explore this effect, we fabricate a $p$-GeS/$n$-MoS$_2$ vdW heterojunction and investigate its electrically and optically tunable anisotropic diode characteristics. The $I$-$V$ response exhibits higher current density and rectification along the armchair direction, with anti-ambipolar transfer characteristics in a back-gated configuration confirming a type-II band alignment and strong gate tunability. Under illumination, the diode shows pronounced anisotropic photoconductance, with time-resolved photocurrent mapping revealing a direction-sensitive response to light stimulation. These findings provide insight into utilizing anisotropic charge transport for future polarization-sensitive optoelectronic and RF rectification applications.

\section*{Experimental details}
\label{subsec:exp}
{\bfseries Synthesis and characterization:} GeS single crystals were synthesized using chemical vapor transport (CVT) method. Ge (99.95\% , Alfa Aesar) and S (99.99\% , Alfa Aesar) powders were mixed thoroughly in stoichiometric ratio and Iodine (4 mg/cc) was added as transport agent. The mixture was loaded into a quartz tube which was then evacuated and sealed under a vacuum of 10$^{-5}$ Torr. The sealed tube was placed in a horizontal two-zone tube furnace and the temperature was slowly increased to 873 K and 773 K for source and growth zone, respectively. The furnace was kept at this temperature for 100 hours followed by natural cooling to room temperature. The quartz ampoule was then cracked open in ambient condition to collect the as-grown GeS single crystals (Fig. S1(a)).

Single crystal X-ray diffraction (SXRD) was carried out using Bruker APEX II (CCD area detector , Mo K$\alpha$ , $\lambda = 0.7107$ {\AA}) at room temperature and the structure solution and refinement was done using the corresponding software packages of the diffractometer. The crystalline phase of the synthesized material was confirmed by room temperature powder X-ray diffraction (XRD Bruker) using Cu-K$\alpha$ radiation followed by the refinement using the Fullprof software. Atomic force microscope (AFM) and Kelvin probe force microscope (KPFM) (Asylum Research MFP-3D) were used to map the thickness and surface potential of the flakes. Lattice planes were characterized using transmission electron microscope (TEM) in a JEOL-JEM-F200 electron microscope by giving an accelerating voltage 200 kV. Elemental analysis was investigated in a scanning electron microscope (SEM: JEOL JSM-6010LA) by energy-dispersive spectroscopic (EDS) study.

For diode measurement, MoS$_2$ flakes were grown on a SiO$_2$ (285 nm)/Si substrate using chemical vapor deposition (CVD) method. An alumina boat containing the substrate and molybdenum oxide (MoO$_3$) powder (2 mg) is placed in the center of a single zone furnace, while sulfur (S) powder (200 mg) was placed 23 cm away from the center at upstream side. Then, after flushing the furnace tube using 200 sccm (standard cubic centimeters per minute) Argon gas for 15 minutes, gas flow was reduced to 20 sccm and the furnace temperature was raised to 800\textdegree C. Finally, after 15 minutes of growth time, the furnace was cooled down naturally, and the substrates were taken out. 

{\bfseries Angle-resolved photoemission spectroscopy:} ARPES measurements were carried out on a freshly cleaved GeS crystal in ultrahigh vacuum at the spectromicroscopy beamline of the Elettra Sincrotrone \cite{arpes1} in Trieste, Italy. The linearly polarized incident radiation was focused through a Schwarzschild objective in order to obtain a spot size of approximately 1 \textmu m. The spectra were acquired at 93 K with two different photon energies of 27 eV and 74 eV. The scanning stage of the sample was employed for positioning and raster imaging the sample with respect to the fixed photon beam. Photoemission intensity maps were taken by rotating the hemispherical analyzer mounted on a two-axis goniometer. The total energy and angular resolutions were better than 50 meV and 0.35\textdegree, respectively.

{\bfseries Angle-resolved polarization Raman spectroscopy:} Raman measurements were performed in back-scattering geometry using a Horiba T64000 Raman spectrometer system equipped with two excitation lasers, a DPSS laser and a He-Ne laser of wavelength 532 nm and 633 nm respectively. The laser beam was focused to approximately 1\textmu m spot size on the sample by using a 50X short-distance (NA= 0.75) microscope objective lens. The laser power was kept below 0.2 mW to avoid damaging the sample. For the polarization-dependence Raman measurements, a half-wave plate was inserted in the path of incident light, and an analyzer was placed before the detector (see Fig. S4(a)). To obtain the parallel polarization configuration, the half wave plate was rotated accordingly such that, the polarization direction of incident and detected beam becomes parallel. The sample was mounted on a custom-made rotation stage, and polarized Raman spectra with different angles were obtained by varying the rotation angle of the sample with a step of 15\textdegree. 

{\bfseries Device fabrication:} For GeS/MoS$_{2}$ heterostructure formation, GeS flakes were transferred onto CVD grown MoS$_2$ using PDMS assisted dry transfer technique. For device fabrication on GeS, MoS$_2$ and heterostructure, standard electron beam lithography was employed to fabricate cross-shaped field-effect transistor devices.  Metal contacts with Au (60 nm) for GeS and Ti/Au (8 nm /60 nm) for MoS$_2$ were deposited by thermal evaporation. FET measurements were carried out using a probe station along with Keithley 2450 and 2601B source measure units (SMU). Photoresponse study of the diode was performed using a 532 nm diode Laser Source, and power density is measured using a Holmarc optical power meter.

\section*{Computational details}
\label{subsec:theory}
First principles density functional theory (DFT) calculations were performed using Vienna Ab initio Simulation Package (VASP) \cite{VASP1,VASP2} to study the structural, electronic and optical properties of bulk GeS. Projector-augmented-wave (PAW) pseudopotential was employed to describe electron-ion interaction. Generalized gradient approximation (GGA) with the Perdew-Burke-Ernzerhof (PBE) \cite{PBE} functional was considered to incorporate the exchange-correlation effects. The band gap underestimation by the traditional PBE method in 2D materials lead us to employ the Heyd–Scuseria–Ernzerhof (HSE06) hybrid functional for a more accurate determination of the band gap of GeS. The lattice parameters and the atomic coordinates were optimized until the Hellmann-Feynman residual forces were less than 0.001 eV/{\AA} per atom, and total energy difference criterion for convergence of electronic self-consistent calculations was set to 10$^{-6}$ eV. The cutoff energy for the plane-wave expansion was set to be 600 eV with a $4\times 10\times 8$ k-point mesh generated according to the Monkhorst-Pack scheme \cite{MP-scheme}. To incorporate the optical anisotropy, absorption spectra were simulated in DFT framework implementing VASPKIT code \cite{VASPKIT}. To visualize the crystal structures VESTA software \cite{VESTA} was used.
For GeS/MoS$_2$ heterostructure the plane wave cut-off energy was set to be 600 eV with a $6\times 6\times 1$ k-point mesh. A vacuum layer of 15 {\AA} was employed to eliminate spurious interactions between adjacent layers.

\section*{Results and discussion}
\label{sec:results_discussions}

\begin{figure}
\centerline{\includegraphics[scale=0.6, clip]{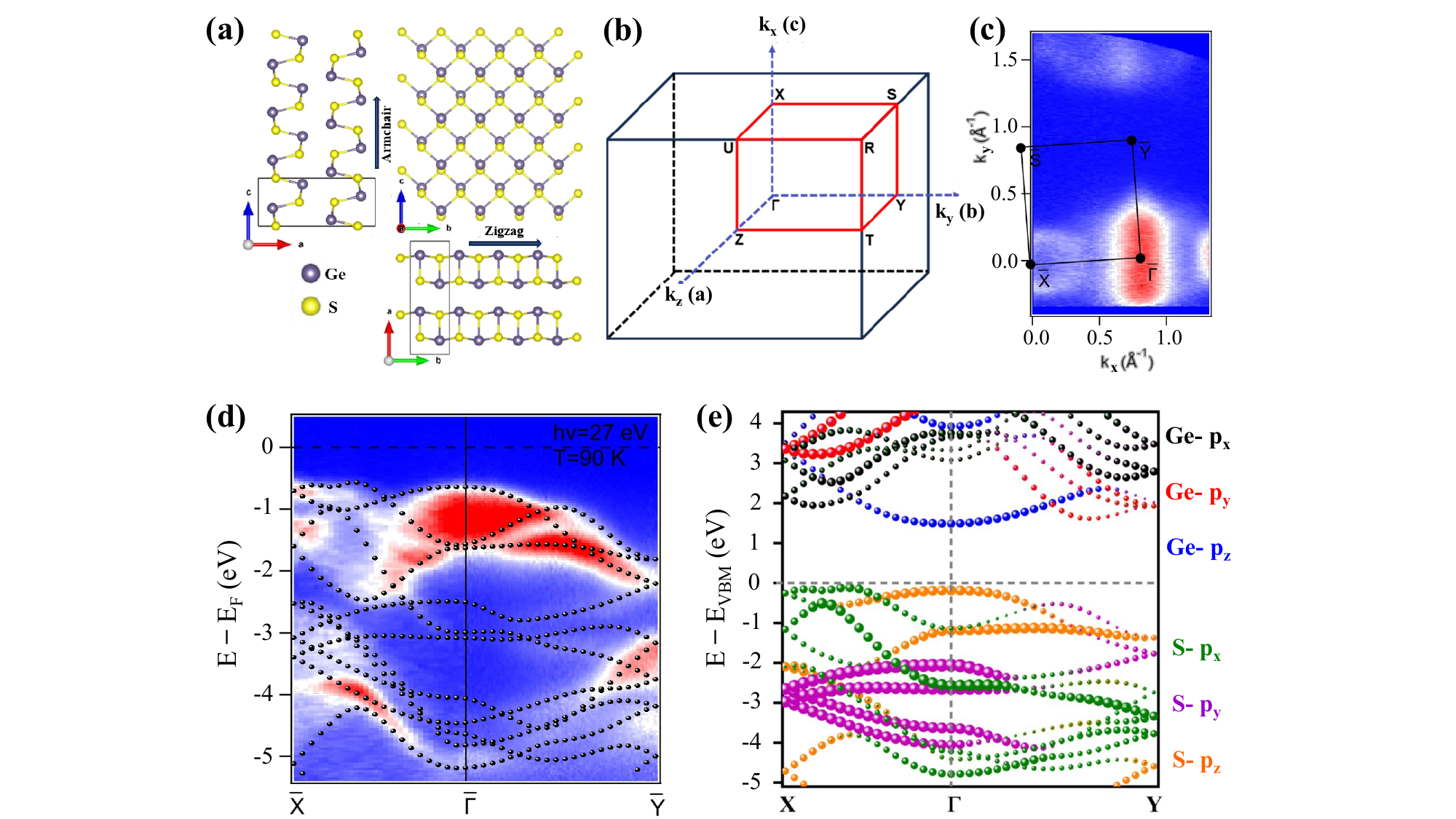}}
\caption{(a) Top and side view of the puckered crystal structure of layered GeS showing the armchair and zigzag chains. The unit cell is denoted by the black rectangular box. (b) Schematic of the 3D orthorhombic Brillouin zone represented with the high symmetry points. (c) Constant energy contour at a binding energy of $−$1 eV. Rectangular symmetry is consistent with the orthorhombic crystal structure. (d) ARPES spectra of $in \ situ$ cleaved GeS crystal at 93 K using 27 eV photon energy and a comparison with the DFT- calculated band structure (black dotted curve) showing anisotropic band dispersion along $\bar{\Gamma}$-$\bar{X}$ and $\bar{\Gamma}$-$\bar{Y}$ direction. (e) Orbital projected band structure calculated using DFT framework.
\label{char}}
\end{figure}

As shown in \fig{char}(a) GeS crystallizes in a 2D-layered structure with orthorhombic space group Pnma (No. 62), with lattice parameters $a=10.494$ {\AA}, $b=3.642$ {\AA}, $c=4.305$ {\AA} and $\alpha=\beta=\gamma=90^{\circ}$ as obtained from SXRD refinement. In a single layer of GeS, Ge atoms are covalently bonded with three adjacent S atoms to form a puckered honeycomb structure, and the atomic layers are stacked together by vdW interactions. Anisotropy in GeS is conceived from this puckered structure of GeS with armchair (AC) and zigzag (ZZ) structural arrangements along in-plane lattice directions $c$ and $b$, respectively.  Powder XRD (Fig. S1(b)) and EDS analysis (Fig. S1(f)) confirmed the phase and stoichiometry (49.93\% of Ge and 50.07\% of S) of the as-grown crystals. GeS exhibits four distinct Raman modes (3$A_g$ and $B_{1g}$) as shown in Fig. S1(c). The AFM mapping and corresponding height profile in Fig. S1(d) determined the thickness of the exfoliated GeS flakes $\sim$50 nm. The lattice planes were characterized by TEM as shown in Fig. S1(e). The d-spacings were calculated to be 2.84 \AA, 2.84 \AA $ $ and 2.16 \AA $ $ for planes (400), (011) and (311), respectively which are consistent with that obtained from the crystallographic information file (CIF) extracted from SXRD refinement. Inset of Fig. S1(e) displays the selected area electron diffraction (SAED) pattern where the first order bright spots denote different Bragg planes. 
\newline

\begin{center}
\bfseries{A. Angle-resolved photoemission spectroscopy and density functional theory}\\
\end{center}

To investigate the electronic structure of GeS, ARPES was performed with a \textmu-focused photon beam using two different photon energies. \fig{char}(b) shows the orthorhombic Brillouin zone (BZ) of GeS, where in-plane lattice axes $b$ and $c$ are considered along $k_y$ and $k_x$ directions, respectively and out-of-plane $a$- axis is taken along $k_z$ direction. ARPES spectra were taken at two different photon energies to examine possible $k_z$ dispersion effects and to aid in identifying the $k_x-k_y$ surface projection of the 3D BZ. Constant energy contour at a binding energy $-$1 eV (\fig{char}(c)) shows surface-projected rectangular symmetry consistent with an orthorhombic crystal structure. As seen from \fig{char}(c), the intense spot-like feature from the spectral weight of valence bands appears around the $\Gamma$ point as observed from the band structure obtained from ARPES using 27 eV photon energy (\fig{char}(d)). Additional spot-like feature appears along $\Gamma$-$X$ direction in the constant energy map which corresponds to another two hole valleys constituting the VBM as observed from band structure. Parabolic VBM towards $X$ direction along with the almost degenerate hole valley at $\Gamma$ point as shown in \fig{char}(d) are situated around a binding energy $\sim-$0.5 eV.
Conversely, along $\Gamma$-$Y$ direction only one valley is observed around $\sim$ 0.5 eV away from VBM. Hole-like valence band is the sharpest using 27 eV photon energy. On the other hand, for 74 eV energy of incident photons the spectrum is much weaker and broader (Fig. S2(b)). In contrast to 27 eV energy, some bands at deeper binding energy appear brighter due to the variation of photoemission cross section of the valence bands. Also, the spectra for 74 eV are slightly shifted upward near the Fermi level because of the out-of-plane dispersion of the bands. 

Considering that the full energy gap between VBM and CBM is 1.6 eV \cite{GeS-opt}, the ARPES result is consistent with the $p$- type behavior of bulk GeS. The experimental valence bands agree well with the calculated band structure, as depicted from the overlay plots in \fig{char}(d). \fig{char} (e) represents the orbital-projected band structure calculated using DFT (the Fermi level is set to the VBM), where the valence bands near the Fermi level are mostly contributed by the $3p$ orbitals of sulfur and the conduction bands are generated due to Ge-$ \ 4p$ orbitals. The VBM towards $\Gamma$-$X$ direction and the nearly degenerate VBM at $\Gamma$ point are entirely contributed by S-$ \ p_x$ and $\ p_z$ orbitals, respectively. Another hole valley along $\Gamma$-$Y$ is conceived from S-$ \ p_y$ orbital. An indirect band gap of $\sim$ 1.5 eV is estimated using HSE06 functional with the CBM situated at $\Gamma$ point. The hole valley at $\Gamma$ point being almost degenerate to VBM, corresponds to a nearly equal direct band gap as well. (The band gap calculated using GGA-PBE functional is $\sim$ 1.2 eV as shown in Fig. S2(c), underestimated because the GGA-PBE functional lacks the derivative discontinuity of the exchange-correlation potential, which is crucial for accurately describing the difference in energy between occupied and unoccupied states. \cite{underestimated gap-1,underestimated gap-2}).

In particular, the most important aspect is that the band structures differ much along two perpendicular in-plane directions from the central $\Gamma$ point evident from both ARPES and DFT. Henceforth, band dispersion is highly anisotropic which affects their electrical transport as well as optical response. Hole-like valence bands are energetically much closer to the Fermi level along $\Gamma$-$X$ direction in which there is armchair-like atomic arrangements in real space.
This anisotropy in the band structure originates from the underlying orthorhombic crystal structure of GeS, which naturally gives rise to distinct in-plane AC and ZZ directions. The puckered arrangement of Ge and S atoms favors stronger orbital hybridization between Ge-$p$ and S-$p$ states along the armchair direction facilitating greater electron delocalization, which in reciprocal space manifests as strongly dispersive electronic bands and thus lower carrier effective masses. In contrast, the zigzag direction supports less efficient orbital overlap, leading to flatter electronic bands and heavier carriers. Thus, the structural anisotropy of GeS directly imprints on its electronic band structure, consistent with our ARPES observations of pronounced band dispersion along the armchair axis.
This may offer electric field driven hole conduction dominated over the zigzag orientation consistent with our transport measurements. 
The hole effective masses ($m_h{^*}$) were calculated from the experimental band structure by fitting the band dispersions with a parabolic function at the valance band maxima. Fig S3(a) and (b) shows the valance band maxima for AC ($k_x$) and ZZ ($k_y$) directions respectively and the solid green lines indicate the fitted curves. Obtained effective masses along the AC and ZZ direction are, 0.5 $m_e$ and 1.1 $m_e$ respectively where $m_e$ is the free electron mass. That is, the effective mass along $k_x$ direction is about 2.1 times lighter than that of the $k_y$ direction.
The effective masses ($m_h{^*}$) were also calculated from DFT to be 0.42 $m_e$ and 0.83 $m_e$ along AC and ZZ directions respectively. The results are consistent with the more dispersive bands along AC direction resulting in higher curvature, thereby lower effective mass. Lower effective mass corresponds to higher mobility as well, corroborated with the field-effect transport properties. Anisotropic optical absorption spectra were also calculated using the dielectric constants obtained along AC and ZZ directions (see supplementary section).

\begin{center}
\bfseries{B. Angle-resolved polarization Raman spectroscopy}\\
\end{center}

The unit cell of GeS (point group: $D2h$) consists of eight atoms which results in a total 24 vibrational branches. According to the group theory analysis, the irreducible zone-center phonon modes can be expressed as $\Gamma$ = $4A_g + 2A_u + 2B_{1g} + 4B_{1u} + 4B_{2g} + 2B_{2u} + 2B_{3g} + 4B_{3u}$. It exhibits 21 optical modes, out of them 2 are inactive, 12 are Raman active (4$A_g$, 2$B_{1g}$, 4$B_{2g}$, and 2$B_{3g}$), and 7 are infrared active (3$B_{1u}$, $B_{2u}$, and 3$B_{3u}$) modes. In the back-scattering Raman geometry, six modes (4$A_g$ mode and 2$B_{3g}$ mode) can be detected when crystallographic $a$ axis is parallel to the propagating vector of the the laser \cite{jpclett-2023,nanomaterials-2021}.

We obtained four characteristic Raman modes at 113 cm$^{-1}$, 214 cm$^{-1}$, 239 cm$^{-1}$, and 270 cm$^{-1}$ excluding the two below 100 cm$^{-1}$ (Fig. S1(c)). The experimental setup for the polarization dependent Raman measurements in back-scattered geometry is schematically shown in Fig. S4(a) and S4(b). In this configuration, the polarization direction of the incident light is described by $\widehat{e}_i$ = (0 $cos\theta$ $sin\theta$) with respect to the crystallographic axes as shown in \fig{char}(a), while for scattered light it is written as, $\widehat{e}_s$ = (0 $cos\theta$ $sin\theta$) and $\widehat{e}_s$ = (0 $sin\theta$ $cos\theta$) for the parallel and cross-polarization configurations, respectively \cite{jpcl-2021,nanoscaleBP-2015}. But, cross polarization doesn't give any extra information, so we have considered only parallel configuration \cite{acsnanoBP-2015,nanolett_phononBP-2016}.

Now, the Raman scattering intensities can be expressed in terms of Raman tensor  $\overleftrightarrow{R_k}$ as, 
\begin{equation}
I_k = |\widehat{e}_i.\overleftrightarrow{R_k}.\widehat{e}_s|^2
\end{equation}
The Raman tensor elements of $\overleftrightarrow{R_k}$ are dependent on the dielectric susceptibility of the material, which has both real and imaginary parts in case of absorptive materials. Hence, the Raman tensor elements should also have complex values as well \cite{acsnanoBP-2015,ReSe2complextensor,prb1998}. So, the Raman tensors for $A_g$ and $B_{3g}$ modes can be written in complex form as,
\begin{equation}
\overleftrightarrow{R}_{A_g} = 
\begin{pmatrix}
\vert a \vert e^{i\phi_a} & 0 & 0\\
0 & \vert b \vert e^{i\phi_b} & 0\\
0 & 0 & \vert c \vert e^{i\phi_c}
\end{pmatrix}
\end{equation}
and
\begin{equation}
\overleftrightarrow{R}_{B_{3g}} = 
\begin{pmatrix}
0 & 0 & 0\\
0 & 0 & \vert g \vert e^{i\phi_g}\\
0 & \vert g \vert e^{i\phi_g} & 0
\end{pmatrix}
\end{equation}

where $\phi$ is the phase factor.
\newline
The corresponding Raman intensities under the parallel polarization configuration are therefore,
\begin{equation}
I^{\|}_{A_g} \propto \vert c \vert^2[(sin^2\theta+\vert\frac{b}{c}\vert cos\phi_{bc}cos^2\theta)^2+\vert\frac{b}{c}\vert^2sin^2\phi_{bc}cos^4\theta]
\end{equation}
\begin{equation}
I^{\|}_{B_{3g}} \propto \vert g \vert^2 sin^22\theta \label{Bg}
\end{equation}
where $\phi_{bc}$ is the phase difference ($\phi_b-\phi_c$).

\begin{figure}
\centerline{\includegraphics[scale=0.5, clip]{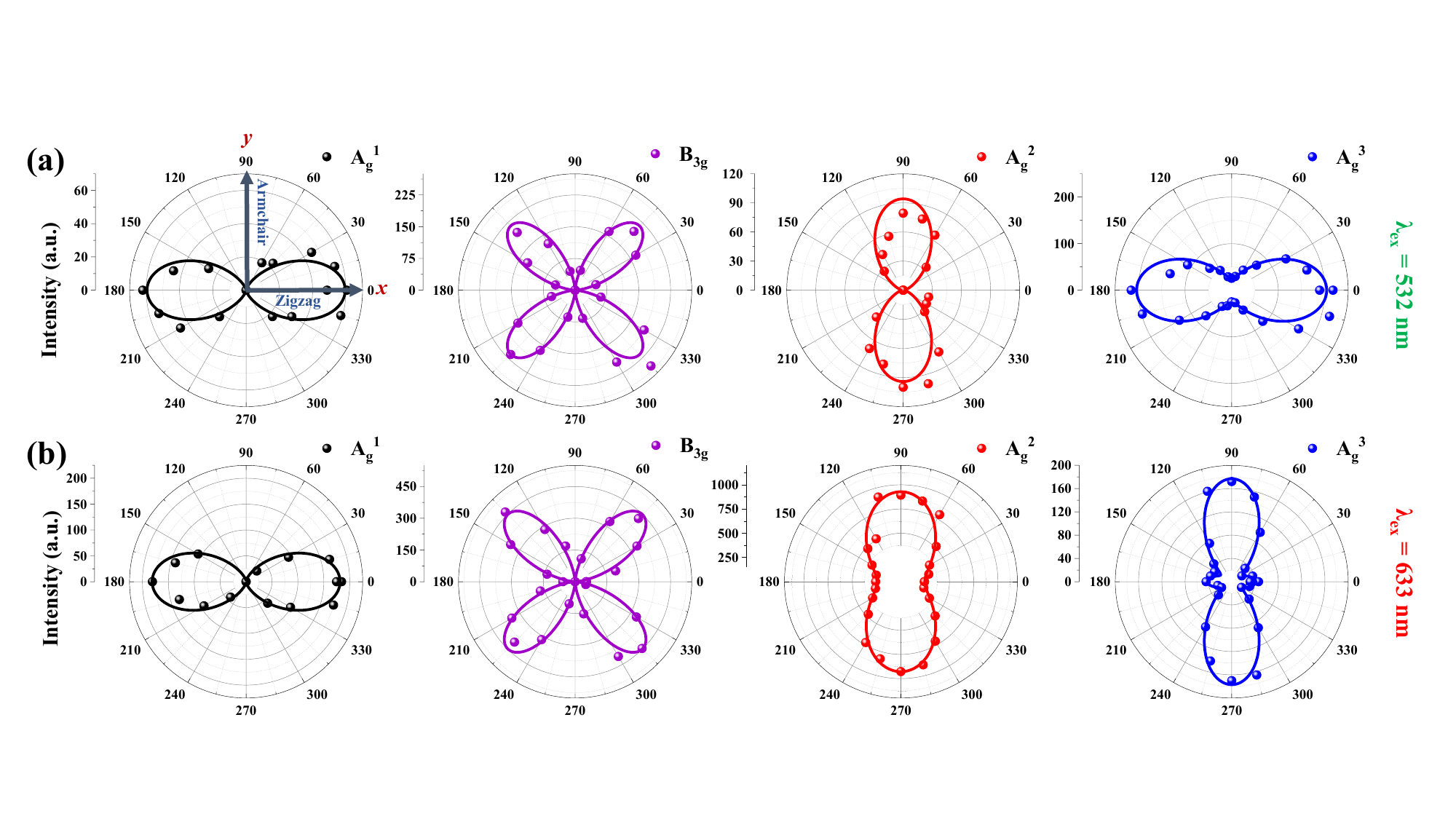}}
\caption{(a) Polar plots of the Raman peak intensities of four GeS phonon modes $A_g^1$, $B_{3g}$, $A_g^2$, and $A_g^3$ measured in parallel polarization configuration under (a) 532 nm and (b) 633 nm excitation. The experimental results are shown in dots, and the solid lines represent fitting curves obtained from the respective intensity equations given in eq. (4) and (5). Grey arrows in the first figure of (a) denote the crystalline orientation (armchair-zigzag) in lab frame (x-y).
\label{Raman}}
\end{figure}

We performed the ARPRS measurements on bulk GeS flake using two excitation wavelengths of 532 nm and 633 nm. The intensity of four detected Raman modes were plotted against the angle of rotation ($\theta$) as illustrated in \fig{Raman}. This angle was primarily set as 0\textdegree$ $ along the $x$ direction of the flake in lab frame, schematically shown in Fig. S4(b). Different Raman modes reveal different angular dependence which is a direct evidence of the in-plane anisotropy of GeS. The Raman intensity of $B_{3g}$ mode exhibits a periodicity of 90\textdegree$ $ with the maximum intensity along both $\theta$ = 0\textdegree$ $ and 90\textdegree$ $. Also, the behavior of this mode remains same for both the excitation lasers, because $B_{3g}$ mode does not depend on the phase factor $\phi$ (see eq. \ref{Bg}). On the other hand, all of the three $A_g$ modes show angular periodicity of 180\textdegree. Intensity of $A_g^1$ mode reaches the maximum value for $\theta$ = 0\textdegree$ $ and is completely undetectable at $\theta$ = 90\textdegree. $A_g^2$ mode possesses two fold symmetry as well, however, in contrast to the $A_g^1$ mode, the direction of maximum intensity is rotated by 90\textdegree. This is because, in $\overleftrightarrow{R}_{A_g}$, $b$ is greater (smaller) than $c$ for $A_g^1$ ($A_g^2$) mode as obtained from the fitting results shown in Table I. The behavior of $A_g^2$ mode gets slightly changed for excitation wavelength 633 nm, where a non zero intensity is observed at the minima \fig{Raman}(b). A small change in the $b/c$ ratio and the contribution of a minimal phase factor $\phi$ were noted for the 633 nm excitation by fitting the data with the intensity equations (see Table I). On the other hand, the dependence of the $A_g^3$ mode on the excitation wavelength is substantially larger. There is a 90\textdegree shift in the direction of maximum intensity for this mode between the two excitations. Two tiny local maxima also pop up along $\theta$ = 0\textdegree$ $ for 633 nm. The observed anomalous behavior can be attributed to a significant influence from $\phi$. Since, the behavior of $A_g^3$ mode is completely different under different laser excitation, whereas $A_g^1$ and $A_g^2$ have no such change, the determination of AC or ZZ direction can be possibly done by comparing the intensities of $A_g^1$ and $A_g^2$ modes only.

H. B. Ribeiro $et \ al.$ \cite{2019PRB} observed in their polarization-dependent Raman experiments that the $A_g^1$ mode has maximum (minimum) intensity along the ZZ (AC) direction. In a similar manner, our 0\textdegree$ $ orientation resembles the crystallographic ZZ direction as shown by the orange arrow in the first figure of \fig{Raman}(a). Our results are also consistent with similar kind of ARPRS studies on GeS conducted by D.Tan $et \ al.$ \cite{2016-nanoresearch}, H. Ouyang $et \ al.$ \cite{Sciencechinamaterial-2020} and Z. Li $et \ al.$ \cite{Acs_applied_materials-2019} as well as for other anisotropic 2D materials with orthorhombic crystal structure \cite{acsnanoBP-2015,nanoscaleSnS-2016}. Related polarization-dependent Raman spectra on exfoliated GeS at the sample's edge showed that either the $A_g^1$ or $A_g^2$ mode is strong along the ZZ or AC direction depending on the respective crystallographic edge and the polarization configuration \cite{2019PRB}.

\begin{center}
\bfseries{C. Field-effect transport measurement}\\
\end{center}

\begin{figure}
\centerline{\includegraphics[scale=0.55, clip]{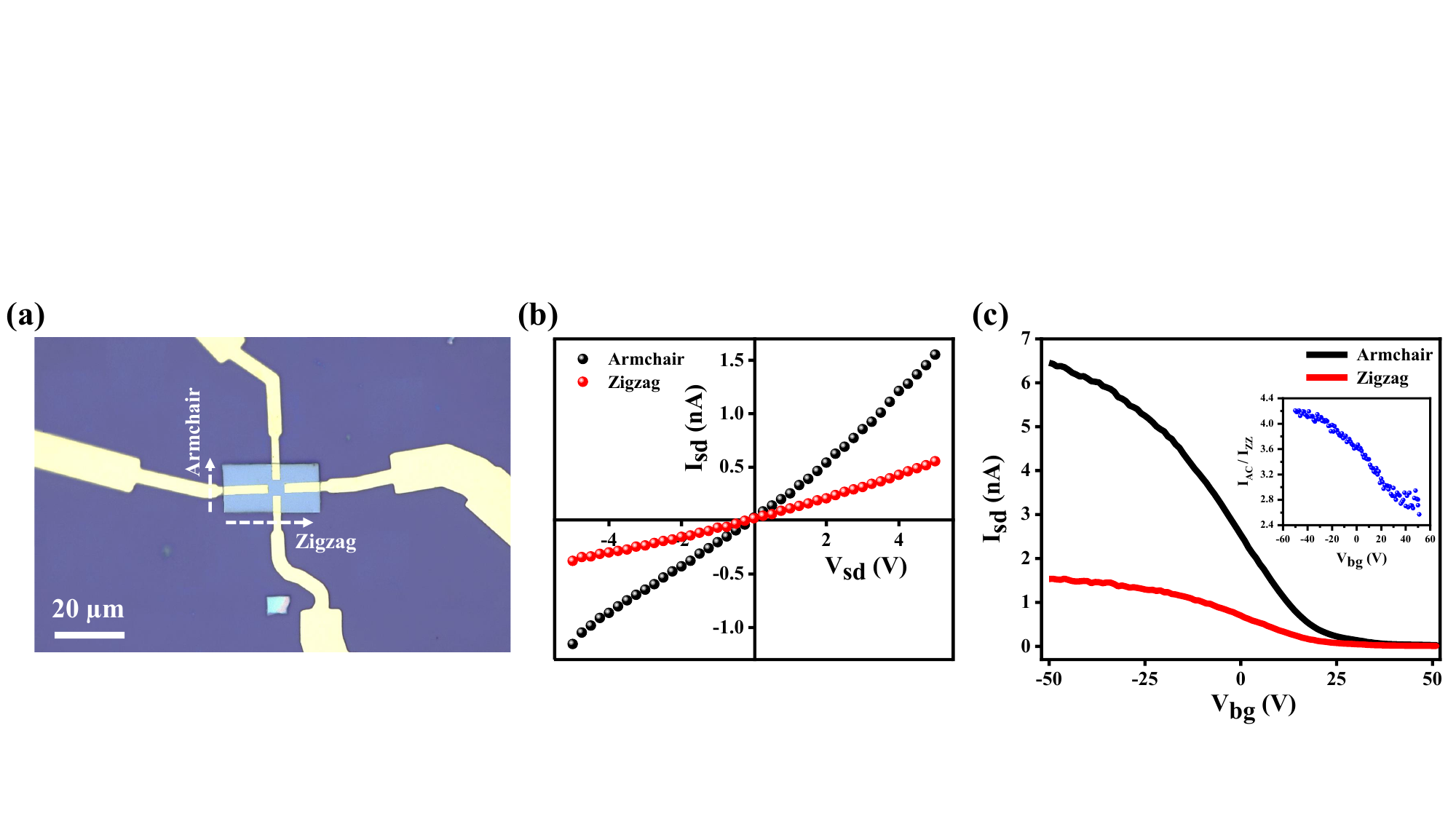}}
\caption{(a) Optical micrograph of a cross-shaped FET device along two orthogonal in-plane orientations of GeS. (b) Two-terminal current-voltage characteristics ($I_{\text{sd}} \ vs. \ V_{\text{sd}}$) of GeS measured along armchair (black) and zigzag (red) direction, indicating higher conductance along armchair orientation. (c) Transfer curve ($I_{\text{sd}} \ vs. \ V_{\text{bg}}$) in back-gated configuration measured along armchair (black) and zigzag (red) direction.
\label{FET}}
\end{figure}

The in-plane anisotropic electrical properties of GeS were investigated through the direction dependent transport measurements. GeS FET devices were fabricated on SiO$_2$/Si substrate using electron beam lithography followed by Au metal electrode deposition by thermal evaporation. As shown in the optical microscope image \fig{FET}(a), four electrodes were fabricated in cross configuration to measure the transport characteristics in two perpendicular directions of multilayered GeS flakes. The AC and ZZ directions were identified by angle-resolved Raman scattering experiment. The channel size between the electrodes in AC and ZZ direction were designed to be the same. As shown in \fig{FET}(b), $I_{\text{sd}}$ (source-drain Current) $vs. \ V_{\text{sd}}$ (source drain voltage) measurement shows clear directional dependence with current along AC direction being much higher than ZZ direction. Also the $I_{\text{sd}}-V_{\text{sd}}$ plots up to $V_{\text{sd}} =$ 5 V reveals nearly linear behavior indicating good ohmic contact with the electrodes. The variation of the source-drain current ($I_{\text{sd}}$) as a function of back-gate voltage ($V_{\text{bg}}$) with 3 V bias ($V_{\text{sd}}$) was measured along both directions. As shown in \fig{FET}(c), $I_{\text{sd}}$ increases with decreasing $V_{\text{bg}}$ from $+$50 V to $-$50 V indicating the typical $p$- type behavior of GeS. This $p$- type nature is attributed to the presence of Ge vacancies in the crystals or exfoliated flakes resulting in high lying valance band of GeS \cite{JAC-2022,inorganicmaterials-2000}.
Furthermore, the current differs between the armchair and zigzag directions, reflecting the inherent anisotropic transport properties of GeS. The AC direction exhibits a higher current than the ZZ direction due to its relatively lower effective mass and higher carrier mobility. Additionally, the ratio of current between the two directions ($I_{\text{AC}}/I_{\text{ZZ}}$) plotted against gate bias (shown in inset of \fig{FET}(c)) slightly increases with decreasing $V_{\text{bg}}$, suggesting that gate modulation enhances the anisotropic transport behavior.  
The two-probe field-effect mobility ($\mu_{\text{eff}}$) was calculated from the linear region of the transfer characteristics using the formula,
\begin{equation}
\mu_{\text{eff,2p}}=\frac{L}{W C_{\text{ox}}} \frac{dI_{\text{sd}}}{dV_{\text{bg}}} \frac{1}{V_{\text{sd}}}
\end{equation}
where \textit{L} and \textit{W} denote the channel length and channel width of the FET device, $C_{\text{ox}}$ is the gate capacitance of the SiO$_2$ layer which can be estimated as,
$C_{\text{ox}}=\frac {\epsilon_0 \epsilon_r}{d}$, 
\textit{d} being the thickness of the SiO$_2$ layer and the corresponding dielectric constant $\epsilon_r$ = 3.9. The carrier mobilities along AC and ZZ direction are calculated to be 0.038 cm$^2$V$^{-1}$s$^{-1}$ and 0.011 cm$^2$V$^{-1}$s$^{-1}$, respectively. These results highlight the strong anisotropy in charge transport within GeS, where the AC direction exhibits significantly higher ($\sim$ 3.4 times) carrier mobility than the ZZ direction.
We have compared the calculated mobilities and ratio with those reported for other TMDCs as well as anisotropic 2D materials in supplementary Table II.
We also measured three other GeS devices along AC and ZZ directions. The extracted mobilities along two directions and their ratios are consistent across devices, confirming the consistency of the observed anisotropy. A summary of the measured values is provided in the supplementary section (Table~III).
This anisotropic behavior is expected to influence the electronic properties of GeS-based heterostructures, including its junction with MoS$_2$, which we explore in the next section.

\begin{center}
\bfseries{D. Anisotropic $p$-$n$ heterojunction diode}\\
\end{center}
To study the effect of anisotropic carrier transport of GeS in a $p$–$n$ junction diode, MoS$_2$ was chosen as $n$-type material due to its high mobility and suitable band gap \cite{PhysRevLettMoS2-2010}. MoS$_2$ flakes were grown on SiO$_2$/Si substrate using the CVD method (Fig.~S6(a)). Raman spectroscopy (Fig.~S6(b)) revealed the two characteristic peaks of MoS$_2$ at 384.1~cm$^{-1}$ and 402.3~cm$^{-1}$ with a separation of 18.2~cm$^{-1}$, indicative of the monolayer nature of the as-grown triangular flakes. The thickness was further confirmed by AFM measurements which showed a step height of $\sim$0.8~nm, and PL spectroscopy displayed the characteristic sharp emission peak at 1.82 eV (Fig.~S6(c,d)).
 $I_{\text{sd}}$-$V_{\text{sd}}$ characteristics of MoS$_2$ (Fig. S6(e)) flakes shows linear behavior indicating ohmic contact with titanium. Transfer curve (Fig. S6(f)) exhibits typical $n$ type behavior of MoS$_2$ with a high on-off ratio of 10$^5$.

\begin{figure}
\centerline{\includegraphics[scale=0.6, clip]{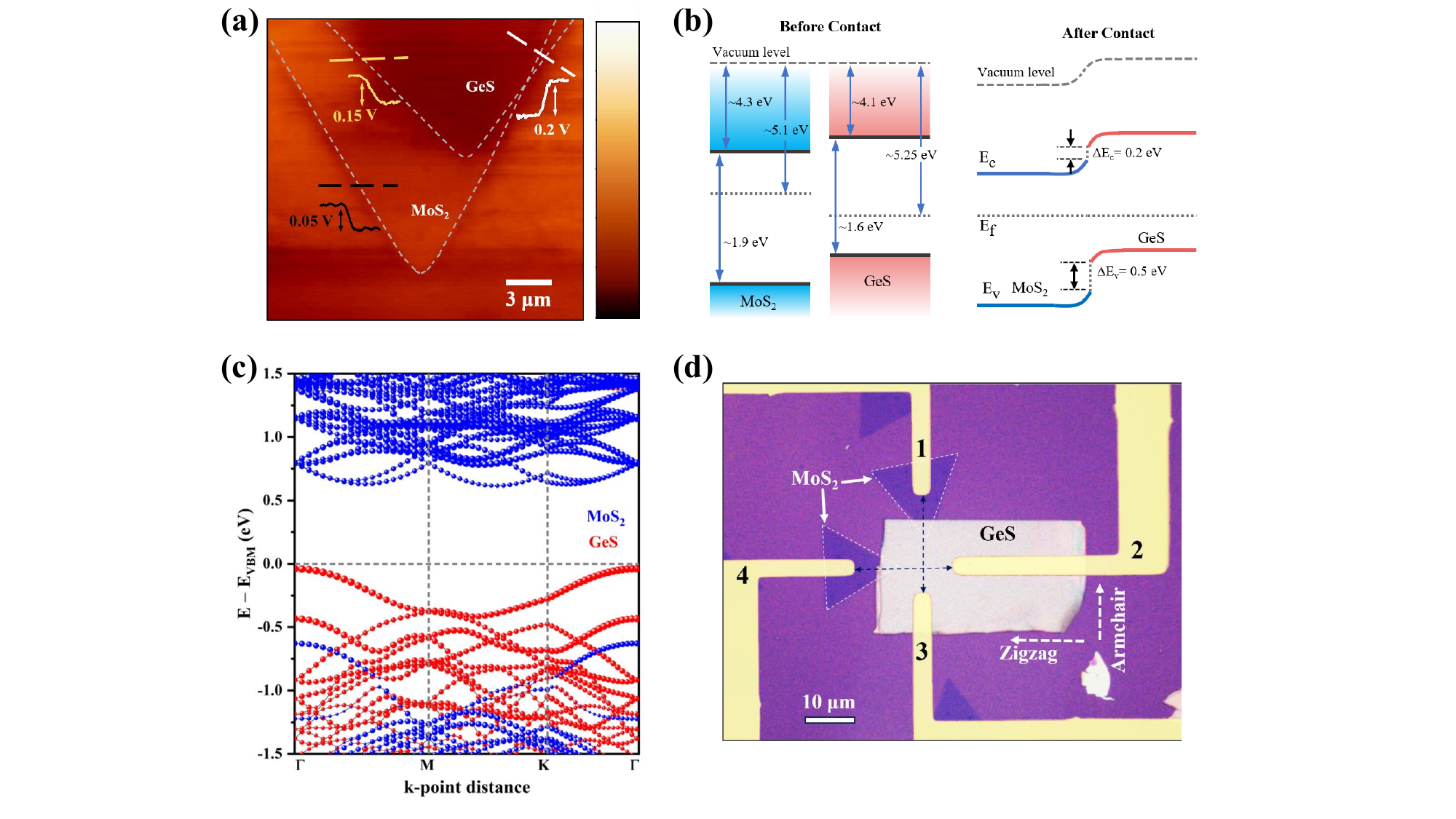}}
\caption{(a) KPFM mapping of the surface potential in GeS (white line-cut), MoS$_2$ (black line-cut) and overlapping region of the heterostructure (yellow line-cut) to obtain the corresponding work functions. (b)  Schematic energy band diagram of GeS/MoS$_2$ heterojunction before (left) and after (right) contact, showing a type-II band alignment across the heterojunction. (c) Layer-projected band structure of GeS/MoS$_2$ heterostructure with VBM contributed by GeS and CBM contributed by MoS$_2$, consistent with the type-II band alignment. (d) Optical microscope image of a cross-shaped device of GeS/MoS$_2$ $p$-$n$ diode including metal contacts along two orthogonal orientations of GeS. Distance between the two sets of opposite electodes are 19.7$\pm$ 0.1  $\mu$m.
\label{diode}}
\end{figure}

Multilayered GeS flake was then transferred onto CVD-grown MoS$_2$ to form the GeS/MoS$_2$ vertical $p$–$n$ heterojunction. Kelvin probe force microscopy (KPFM) was employed to investigate the surface potential distribution of the GeS/MoS$_2$ heterostructure, providing insights into its built-in electric field and charge transfer behavior. KPFM measures the contact potential difference (CPD) between the conducting AFM tip and the sample surface, which is related to the work function ($\phi$) by, 
\begin{equation}
V_{\text{CPD}} = \frac{1}{e} (\Phi_{tip} - \Phi_{sample})
\end{equation}
where $\Phi_{\text{tip}}$ and $\Phi_{\text{sample}}$ are the work functions of the tip and the sample, respectively.
In the case of a MoS$_2$ flake on a SiO$_2$ substrate, the work function of the MoS$_2$ flake can be calculated from the measured V$_{\text{CPD}}$ difference between MoS$_2$ and SiO$_2$ and reported reference work function of SiO$_2$ ($\Phi_{\text{substrate}}$ = 5.05 eV) \cite{ApplPhysLett-2009}, using the following equation,
\begin{equation}
\Delta V_{\text{CPD}} = V_{\text{CPD}}(\text{sample}) - V_{\text{CPD}}(\text{substrate}) = \frac{1}{e} (\Phi_{\text{substrate}} - \Phi_{\text{sample}})
\end{equation}
From our KPFM micrograph on a GeS/MoS$_2$ heterostructure shown in \fig{diode}(a), the $\Delta V_{\text{CPD}}$ values for MoS$_2/$SiO$_2$, GeS/MoS$_2$ and GeS/SiO$_2$ are found out to be $-$0.05 V, $-$0.15 V, and $-$0.20 V, respectively. 
Using these values in the above equation, the extracted work function values for MoS$_2$ and GeS were found to be 5.1 eV and 5.25 eV, respectively, which align well with reported values \cite{JKPS-2014,CrystalsGeS-2022}.
Based on these measurements, a band alignment diagram for the GeS/MoS$_2$ heterostructure is illustrated (shown in \fig{diode}(b)) using the extracted work function values and previously reported electron affinities, 4.3 eV for MoS$_2$ \cite{NanoLettMoS2-2016} and 4.1 eV for GeS \cite{CrystalsGeS-2022}. The results indicate the formation of a type-II heterostructure, where the CBM and VBM are staggered, facilitating  efficient charge separation. The built-in potential at the interface plays a crucial role in driving carrier transport, which is further modulated by external bias, as observed in electrical and optoelectronic measurements.

The interfacing between GeS and MoS$_2$ layers are studied using density functional theory. The lattice parameter of the combined GeS/MoS$_2$ is calculated to be 8.48 {\AA} with a lattice mismatch 0.6 \% and interlayer distance 3.40 {\AA}. The projected band structure of GeS/MoS$_2$ vdWH is depicted in \fig{diode}(c) using GGA-PBE functional to minimize the computational cost. The vdWH possesses an indirect band gap of 0.6 eV. The VBM lies at the $\Gamma$ point, mostly contributed by GeS and the CBM lies between $M$-$K$ path, mostly comes from the MoS$_2$ layer. The vdWH thus forms a type-II band alignment. 
To further elucidate the electronic nature of the GeS/MoS$_2$ heterostructure, the partial charge densities corresponding to the valence band maximum (VBM) and the conduction band minimum (CBM) are analyzed as shown in Fig. S7(a). The charge density distribution at the VBM is predominantly localized on the GeS layer, while the CBM states are primarily contributed by the MoS$_2$ layer. This clear spatial separation of the frontier states indicates a type-II band alignment in the heterostructure. The charge density difference (CDD) is also analyzed in order to describe the interfacial charge redistribution. The CDD is defined as:
\begin{equation}
\Delta \rho (z) = \rho_{\text{vdWH}} (z) - \rho_{\text{GeS}} (z) - \rho_{\text{MoS$_2$}} (z)
\end{equation}
where the charge densities of vdWH, isolated GeS and MoS$_2$ layers are denoted as $\rho_{\text{vdWH}}$, $\rho_{\text{GeS}}$, and $\rho_{\text{MoS$_2$}}$, respectively. The CDD plotted in Fig. S7(b) represents the positive charge accumulation near GeS layer whereas negative charge depletion near MoS$_2$ layer. This depicts the preferential electron transfer from the MoS$_2$ layer to the GeS layer. Hence, GeS acts as an electron acceptor, while the MoS$_2$ acts as an electron donor, consistent with the expected band alignment.

\begin{figure}
\centerline{\includegraphics[scale=0.6, clip]{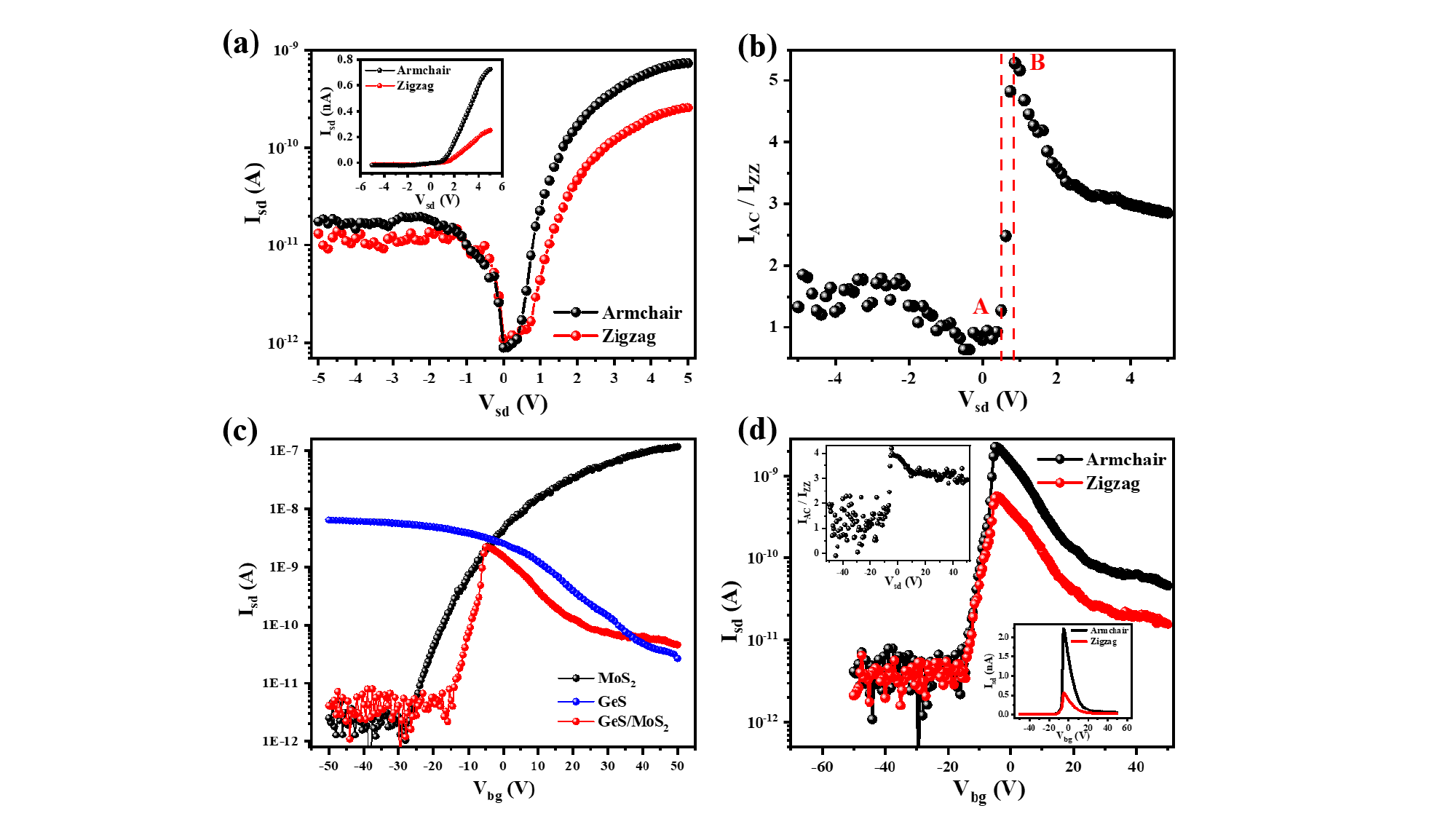}}
\caption{(a) $I_{\text{sd}} \ vs. \ V_{\text{sd}}$ characteristics in semi-log scale measured across the $p$-$n$ heterojunction showing anisotropic response along armchair and zigzag directions of GeS. Inset shows the diode characteristics in linear scale. (b) Anisotropic current ratio of the $p$-$n$ diode plotted against the bias voltage ($V_{\text{sd}}$) indicating two threshold points A and B for armchair and zigzag directions, respectively. (c) Transfer characteristics  of the back-gated heterojunction FET in semi-log scale (red) showing an anti-ambipolar characteristics which is effectively a superposition of the transfer characteristics of $p$-type GeS (blue) and $n$-type MoS$_2$ (black). (d) Anti-ambipolar transfer characteristics of GeS/MoS$_2$ in semi-log scale along two different orientations at $V_{\text{sd}}=$ 3 V. Inset (bottom right) shows the $I_{\text{sd}}-V_{\text{bg}}$ plot in linear scale. Anisotropic current ratio for armchair and zigzag directions is plotted against the back-gate voltage ($V_{\text{bg}}$) in the top left inset.   
\label{diode-FET}}
\end{figure}

The electrical transport properties of GeS/MoS$_2$ heterostructure along two in-plane directions of GeS were systemically studied under atmospheric condition at room temperature. The optical image of the cross-shaped heterostructure device is shown in \fig{diode}(d) where two MoS$_2$ flakes are in contact with two different directions of GeS, each with similar contact area and channel dimensions. The diode characteristics were measured between electrodes 1 and 3, corresponding to charge transport along the AC direction of GeS, and between electrodes 2 and 4, corresponding to the ZZ direction. The $I_{\text{sd}}-V_{\text{sd}}$ characteristics of the heterostructure device measured along AC and ZZ directions of GeS is represented in \fig{diode-FET}(a), where both curves exhibit an asymmetric, diode-like rectifying behavior, with higher current observed under forward bias (when the positive terminal is connected to $p$-type GeS and the negative terminal to $n$-type MoS$_2$). This behavior indicates charge transfer and the formation of a depletion region at the heterostructure interface.
It should be noted, that the built-in potential is primarily determined by the work function difference and carrier concentrations of MoS$_2$ and GeS. Since this parameters do not have any direction dependence, no significant directional variation in barrier height or depletion width is expected.
However, at the forward bias the anisotropy of GeS strongly influences the diode behavior resulting in higher current along the AC direction compared to that along the ZZ direction.
The anisotropic diode current ratio, defined as the ratio of current along the AC direction to that along the ZZ direction, was plotted with respect to bias voltage in \fig{diode-FET}(b), for the GeS/MoS$_2$ heterostructure. The results reveal that the current ratio is significantly higher under forward bias than under reverse bias. This behavior arises because under forward bias, carrier injection and transport are strongly influenced by the higher mobility along the AC direction, leading to enhanced current flow. In reverse bias, however, transport is mainly limited by the depletion region, reducing the influence of anisotropy.
Consequently, the rectification ratio (the ratio of forward to reverse current) is notably higher along the AC direction, calculated to be 41 at 5 V bias, whereas it is 19 along the ZZ direction, further highlighting the impact of anisotropy on the device's performance. Additionally, the anisotropic current ratio versus bias voltage plot reveals distinct threshold voltages for the two directions. At approximately 0.5 V, the current along the AC direction begins to increase significantly, leading to a sharp rise in the ratio. However, near 0.8 V, the current along the ZZ direction also starts to increase, causing the ratio to decline and eventually saturate at a value close to 3 at higher voltages. The difference in threshold voltages can be attributed to the higher carrier mobility and along AC direction, allowing easier carrier injection and charge transport at slightly lower voltage compared to ZZ directions.

The transfer characteristics of the heterostructure device, as shown in \fig{diode-FET}(c) and (d), exhibit distinct anti-ambipolar behavior when measured along both directions at 3 V bias by sweeping the back-gate voltage from 50 V to $-$50 V. The anti-ambipolar transfer characteristics can be qualitatively understood as a superposition of the $p$-type and $n$-type FET transfer curves, as shown in \fig{diode-FET}(c), where their overlapping conduction regimes give rise to the characteristic anti-ambipolar dependence. This behavior generally originates from the unique gate-tunable electronic properties of the heterostructure with a type-II band alignment,where the CBM of the $n$-type material is lower than the VBM of the $p$-type material, facilitating charge separation accross the layers and gate tunable control over both type of carrier transport. Here, the current exhibits a characteristic $\Lambda$-shaped dependence on the gate voltage, where the current is maximum at intermediate gate voltages and diminishes at extreme positive or negative gate biases. The underlying mechanism involves gate-induced shifts in the energy band alignment between the $p$- and $n$-type regions, which regulate charge transport across the junction. At the peak, the Fermi levels align optimally for efficient charge transport, enabling significant injection of both electrons and holes across the junction. In contrast, at extreme gate voltages, one side becomes heavily doped while the other becomes depleted, creating a high energy barrier that suppresses carrier transport \cite{AdvFunctMaterAntiambipolar-2020,AdvFunctMater_2024_doublevdW}.
Additionally, the transfer curves along the AC and ZZ directions (shown in \fig{diode-FET}(d)) display significant anisotropy on the positive gate voltage ($V_{\text{bg}}$) side of the peak, whereas on the negative $V_{\text{bg}}$ side, the anisotropy  diminishes sharply. This behavior is also reflected in the anisotropic current ratio plotted as a function of gate voltage, as shown in inset (top left) of \fig{diode-FET}(d). The observed anisotropy on the positive $V_{\text{bg}}$ side arises because, in this region, the current is primarily determined by the hole-dominated GeS, which exhibits notable directional dependence in its electronic properties. Consequently, the current varies depending on whether the measurement is made along the AC or ZZ direction. On the other hand, on the negative gate voltage side, the current is mainly governed by MoS$_2$. As a result, the influence of GeS’s anisotropy is significantly reduced, leading to similar current response for different crystallographic orientations of GeS.

To further assess the anti-ambipolar behavior of the GeS/MoS$_2$ heterostructure, we extracted key parameters from the transfer characteristics. The device exhibits a distinct anti-ambipolar peak with a maximum drain current ($I_{\text{peak}}$) of 2.23 nA in the AC direction and 0.55 nA in the ZZ direction, occurring at a gate voltage ($V_{\text{peak}}$) of $-$4.5 V. The anti-ambipolar behavior is observed over a gate voltage range of approximately $-$15 V to $+$25 V for both directions. We note that the off-state current on the MoS$_2$ side is lower than that on the GeS side. This asymmetry originates from the difference in gate tunability between the two channel materials: monolayer MoS$_2$, being atomically thin, exhibits stronger gate control and therefore achieves a lower off-state current, whereas the thicker GeS flake shows weaker gate modulation, resulting in a higher off-state current. Consequently, the peak-to-valley ratio (PVR), defined as $I_{\text{peak}}/I_{\text{valley}}$, is approximately 25 and 35 on the GeS side for the ZZ and AC directions respectively, and increases to around 250 and 400 on the MoS$_2$ side. Such asymmetry in valley currents has also been reported in other vdW heterostructures and can impact the symmetry and gain characteristics in circuit applications \cite{Device_2024_GaAsSb_MoS2,ACS_AMI_2024,ACSAMI_2024_SnSe2MoTe2}.

We further checked the reproducibility of the anisotropic diode characteristics by  measuring two additional GeS/MoS$_2$ heterostructure diodes. The rectification ratios and anti-ambipolar characteristics are in good agreement with those presented above, confirming the reliability of the anisotropic diode performance. A summary of the results is provided in the supplementary section Table~IV.

\begin{figure}
\centerline{\includegraphics[scale=0.6, clip]{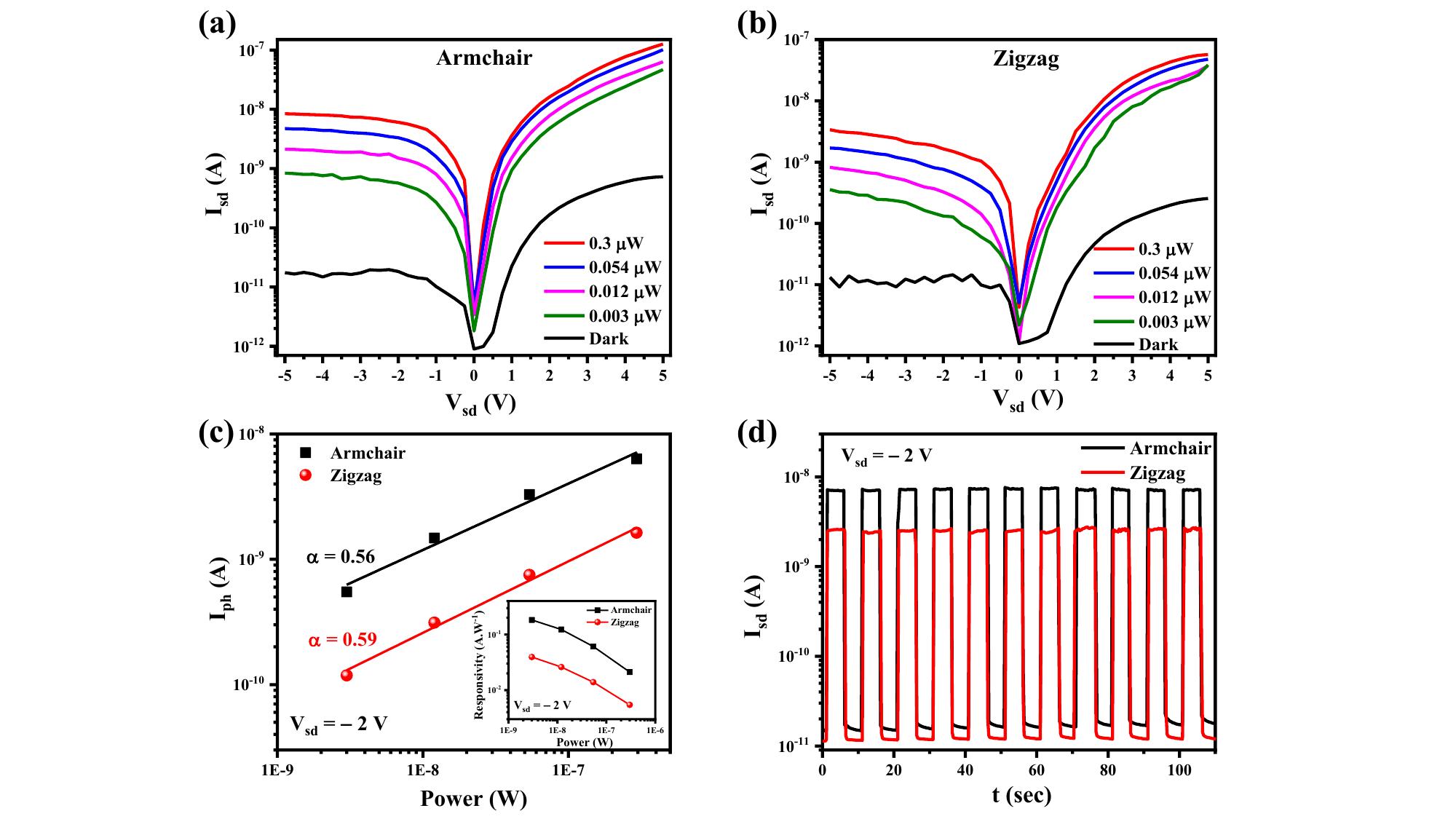}}
\caption{$I_{\text{sd}} \ vs. \ V_{\text{sd}}$ characteristics in semi-log scale measured across the $p$-$n$ heterojunction along (a) armchair, (b) zigzag directions of GeS in dark and illuminated conditions using 532 nm Laser with various power. (c) Photocurrent as a function of Laser power along armchair and zigzag orientations at $-$2 V bias using 532 nm Laser source. Inset shows the responsivity $vs.$ Laser power calculated along armchair and zigzag direction. (d) Photoswitching response of the GeS/MoS$_2$ heterojunction diode at $V_{\text{sd}}= -$2 V. Time-dependent $I_{\text{sd}}$ denoted by black and red color correspond to the armchair and zigzag orientations, respectively.      
\label{photoresponse}}
\end{figure}

The optoelectronic properties of the GeS/MoS$_2$ heterostructure diode were then systematically investigated to evaluate the influence of anisotropy on its photoresponse behavior. Under laser illumination, photogenerated carriers are created and separated at the $p$-$n$ interface of the heterostructure. This separation process is particularly efficient under reverse bias, where the depletion region widens, further enhancing the carrier collection and contributing to a measurable increase in photocurrent. $I$-$V$ characteristics were measured under varying light power levels using 532 nm illumination for both directions of GeS and the results are plotted in semi-log scale, shown in \fig{photoresponse}(a) and (b) for the AC and ZZ respectively. Notably, the photocurrent under reverse bias increased consistently with laser power, and for all power levels, the current along the AC direction of GeS was consistently higher than that along the ZZ direction, emphasizing the strong anisotropic photocurrent transport of the heterostructure. The photocurrent-to-dark-current ratios at $-$5 V reverse bias were observed to be approximately $6\times10^2$ for the AC direction and $2.1\times10^2$ for the ZZ direction, demonstrating superior performance along the AC direction.
The photocurrent ($I_{\text{ph}} = I_{\text{light}} - I_{\text{dark}}$) as a function of optical power ($P_{\text{opt}}$) for both the armchair (AC) and zigzag (ZZ) directions is plotted in double log scale and shown in \fig{photoresponse}(c). The data were fitted using the power$-$law relation $I_{ph} \propto P^\alpha$, yielding $\alpha = 0.56$ (AC) and $\alpha = 0.59$ (ZZ). The sublinear dependence ($\alpha < 1$) indicates that while photocurrent increases with optical power, the rate of increase diminishes at higher illumination levels. 
Further, the photoresponsivity (R) was calculated at different optical power levels. Photoresponsivity, defined as $R = I_{\text{ph}} / P_{\text{opt}}$, measures the efficiency of a photodetector in converting incident light into an electrical signal. Calculated responsivity at $-$2V bias for both direction using different laser power is shown in the inset of \fig{photoresponse}(c). The results  show that the AC and ZZ direction exhibits responsivity of 0.021 A.W$^{-1}$ and 0.0054 A.W$^{-1}$ at an incident power of 0.3 $\mu$W, however the responsivity decreases with increasing power for both directions.
The reduction in responsivity at higher excitation powers can be attributed to the combined effects of trap state saturation and enhanced recombination dynamics. At low power, trap states can temporarily capture carriers and extend their lifetime, thereby contributing to higher photocurrent. However, once these traps become filled under stronger illumination, they lose this role, and the effective carrier lifetime decreases. In addition, the higher photocarrier density promotes non-radiative processes such as Auger recombination, which further limits the number of carriers available for collection \cite{NanoLett_2014_BP,AdvSci_2021_BPGrapheneInSe}. Together, these effects lead to sublinear photocurrent scaling and a decrease in responsivity with increasing optical power.
Also, time-dependent current measurements at $-$2 V bias using a 5-second light pulse revealed a sharp rise and fall in the current for both direction, indicating a rapid response to light stimulation (\fig{photoresponse}(d)). Notably, no photocurrent was observed at zero bias, suggesting that the built-in potential alone is insufficient for efficient carrier separation. This can be explained by a combination of factors, including a high recombination rate of photogenerated carriers and trap-assisted carrier capture, which limit the measurable photocurrent at zero bias highlighting the necessity of external field.

To examine the wavelength dependence of the anisotropy, we have also measured the photoresponse at 633 nm excitation. The corresponding photocurrent $vs.$ power and responsivity $vs.$ power plots are provided in the supplementary section (Fig. S8). Both the photocurrent and responsivity decrease at the longer wavelength. However, the armchair-to-zigzag ratio of photocurrent remains nearly constant, $\sim$3.89 for 532 nm and $\sim$34.24 for 633 nm excitation, demonstrating that the anisotropy does not vary significantly with excitation wavelength.
These findings reinforce the role of anisotropy in efficient charge transport and photocurrent generation in the GeS/MoS$_{2}$ heterostructure.

\section*{Conclusions}
\label{sec:conclusion} 

In conclusion, the intrinsic anisotropy of GeS, revealed through ARPES and Raman spectroscopy, manifests in its electronic and transport properties. Direction-dependent FET measurements confirm anisotropic charge transport with higher hole conductivity along the armchair direction. The GeS/MoS$_2$ $p$-$n$ diode exhibits enhanced rectification and photoresponse along the armchair axis, accompanied by distinct antiambipolar transfer characteristics. These findings provide fundamental insight into anisotropic charge transport and pave the way for GeS-based vdW heterojunctions in polarization-sensitive photodetectors, and multi-valued logic circuits.

\section*{Author contributions}
R. Paramanik grew the single crystals and CVD flakes, fabricated the devices, performed transport and optoelectronic measurements, analyzed data, and contributed to paper writing. T. Kundu conducted computational studies, participated in ARPES experiments, and contributed to data analysis and manuscript preparation. S. Das designed and performed polarized Raman measurements and contributed to data analysis. A. Barinov assisted in ARPES experiments at the beamline. B. Das and B. Karmakar contributed to sample growth and transport measurements. S. Maity and M. Palit carried out electron beam lithography. K. Dolui contributed in the computational calculations. S. K. Mahatha led the ARPES measurements and contributed to data analysis. S. Datta supervised the project, designed the experiments, and contributed to manuscript preparation. 

\section*{Conflicts of interest}
There are no conflicts of interest to declare.

\section*{Data availability}
The data supporting this article have been included as part of the supplementary information.

\begin{acknowledgments}
RP acknowledges the financial support from CSIR, TK and SM acknowledge DST-INSPIRE, SDas is grateful to UGC, and BD, BK, and MP appreciate IACS for the fellowship. RP would like to thank Anupam Pal for his technical support during the growth of the crystal. RP is also thankful to Dr. K.D.M. Rao and Sougata Karmakar for experimental help. CSS facilities of IACS are greatly acknowledged. TK is thankful to IACS HPC cluster facility. SD acknowledges the financial support from DST-SERB grant No. SCP/2022/000411 and CRG/2021/004334. SD also acknowledges the facility e-beam lithography, Technical Research Centre (TRC), IACS, Kolkata. The research leading to \textmu-ARPES in Elettra, Italy under Proposal No. 20230351 supported by a grant from the Italian Ministry of Foreign Affairs and International Cooperation and the Indian Department of Science \& Technology, is greatly acknowledged.  
 
\end{acknowledgments}

\newpage

\end {document}


\title{Supplementary Information:
\\
In-Plane Anisotropy-Driven Directional Charge Transport in van der Waals \textit{p-n} Heterojunction}

\author{Rahul Paramanik}
\affiliation{School of Physical Sciences, Indian Association for the Cultivation of Science, 2A $\&$ B 
Raja S. C. Mullick Road, Jadavpur, Kolkata- 700032, India}

\author{Tanima Kundu}
\affiliation{School of Physical Sciences, Indian Association for the Cultivation of Science, 2A $\&$ B 
Raja S. C. Mullick Road, Jadavpur, Kolkata- 700032, India}

\author{Soumik Das}
\affiliation{School of Physical Sciences, Indian Association for the Cultivation of Science, 2A $\&$ B 
Raja S. C. Mullick Road, Jadavpur, Kolkata- 700032, India}

\author{Alexey Barinov}
\affiliation{Sincrotrone Trieste s.c.p.a., 34149 Basovizza, Trieste, Italy}

\author{Bikash Das}
\affiliation{School of Physical Sciences, Indian Association for the Cultivation of Science, 2A $\&$ B 
Raja S. C. Mullick Road, Jadavpur, Kolkata- 700032, India}

\author{Bipul Karmakar}
\affiliation{School of Physical Sciences, Indian Association for the Cultivation of Science, 2A $\&$ B 
Raja S. C. Mullick Road, Jadavpur, Kolkata- 700032, India}

\author{Sujan Maity}
\affiliation{School of Physical Sciences, Indian Association for the Cultivation of Science, 2A $\&$ B 
Raja S. C. Mullick Road, Jadavpur, Kolkata- 700032, India}

\author{Mainak Palit}
\affiliation{School of Physical Sciences, Indian Association for the Cultivation of Science, 2A $\&$ B 
Raja S. C. Mullick Road, Jadavpur, Kolkata- 700032, India}

\author{Kapildeb Dolui}
\affiliation{Department of Physics, Indian Institute of Technology Tirupati, Tirupati, Andhra Pradesh 517619, India}

\author{Sanjoy Kr Mahatha}
\email{sanjoymahatha@gmail.com} 
\affiliation{UGC-DAE Consortium for Scientific Research, Khandwa Road, Indore 452001, Madhya Pradesh, India}

\author{Subhadeep Datta}
\email{sspsdd@iacs.res.in}
\affiliation{School of Physical Sciences, Indian Association for the Cultivation of Science, 2A $\&$ B 
Raja S. C. Mullick Road, Jadavpur, Kolkata- 700032, India}

\maketitle
\newpage
\section{Raman tensor notations}
Various coordinate notations have been used in the literature. For example, some use $a$ as armchair, $b$ as zigzag, $c$ as out-of-plane directions \cite{natcom-BP,natnanotech-BP}; some use $a$ as armchair, $c$ as zigzag, $b$ as out-of-plane direction \cite{acsnano-2015} or $b$ as armchair, $c$ as zigzag, $a$ as out-of-plane \cite{ssc-1985}. Notably, different notations result in distinct Raman mode representations, especially for $B_g$ mode. For instance, the $B_g$ mode is represented as $B_{1g}$, $B_{2g}$, and $B_{3g}$, respectively, if we choose $c$, $b$, and $a$ as the out-of-plane direction \cite{nanolett-2016}. For Orthorhombic GeS crystallizing in $Pnma$ space group, $a$ is the out-of-plane direction, $b$ is zigzag, and $c$ is armchair direction.
\begin{equation}
\overleftrightarrow{R}_{A_g} = 
\begin{pmatrix}
a & 0 & 0\\
0 & b & 0\\
0 & 0 & c
\end{pmatrix},
\overleftrightarrow{R}_{B_{1g}} = 
\begin{pmatrix}
0 & d & 0\\
d & 0 & 0\\
0 & 0 & 0
\end{pmatrix},
\overleftrightarrow{R}_{B_{2g}} = 
\begin{pmatrix}
0 & 0 & f\\
0 & 0 & 0\\
f & 0 & 0
\end{pmatrix},
\overleftrightarrow{R}_{B_{3g}} = 
\begin{pmatrix}
0 & 0 & 0\\
0 & 0 & g\\
0 & g & 0
\end{pmatrix}
\end{equation}
\newline
\begin{table}[h!]
\renewcommand{\thetable}{S\Roman{table}}
\caption{\label{tab:widgets} Fitting parameters for the polar plots of GeS Raman modes}
.\\
\begin{tabular}{|c|c|c|c|c|c|c|c|c|}
\hline
 Raman&\multicolumn{4}{|c|}{532 nm} &\multicolumn{4}{|c|}{633nm}\\[1.5ex]
\cline{2-9}
Modes&$b$&$c$& $ \ \ b/c \ \ $&$\phi$&$b$&$c$& $ \ \ b/c \ \ $&$\phi$\\[1.5ex]
\hline
$A_g^1$&$7.7\pm0.2$&$0.5\pm0.2$&$15.1\pm5.9$&$0$&$13.5\pm0.3$&$1.6\pm0.8$&$8.3\pm4.3$&$0$\\[1.5ex]
$A_g^2$&$1.4\pm0.7$&$9.7\pm0.3$&$0.1\pm0.1$&$0$&$15.7\pm0.6$&$30.5\pm0.4$&$0.5\pm0.1$&$34.0\pm11.8$\\[1.5ex]

$A_g^3$&$14.3\pm0.2$&$5.2\pm0.6$&$2.7\pm0.3$&$52.0\pm8.1$&$6.4\pm0.2$&$13.3\pm0.1$&$0.5\pm0.1$&$112.0\pm4.3$\\[1.5ex]

\hline \hline
&\multicolumn{4}{|c|}{g} &\multicolumn{4}{|c|}{g}\\[1.5ex]
\hline
$B_{3g}$&\multicolumn{4}{|c|}{$14.7\pm0.2$} &\multicolumn{4}{|c|}{$21.2\pm0.3$}\\[1.5ex]
\hline
\end{tabular}
\\

\end{table}

\section{Absorption Spectroscopy}
\begin{figure}[ht]
\centerline{\includegraphics[scale=0.6, clip]{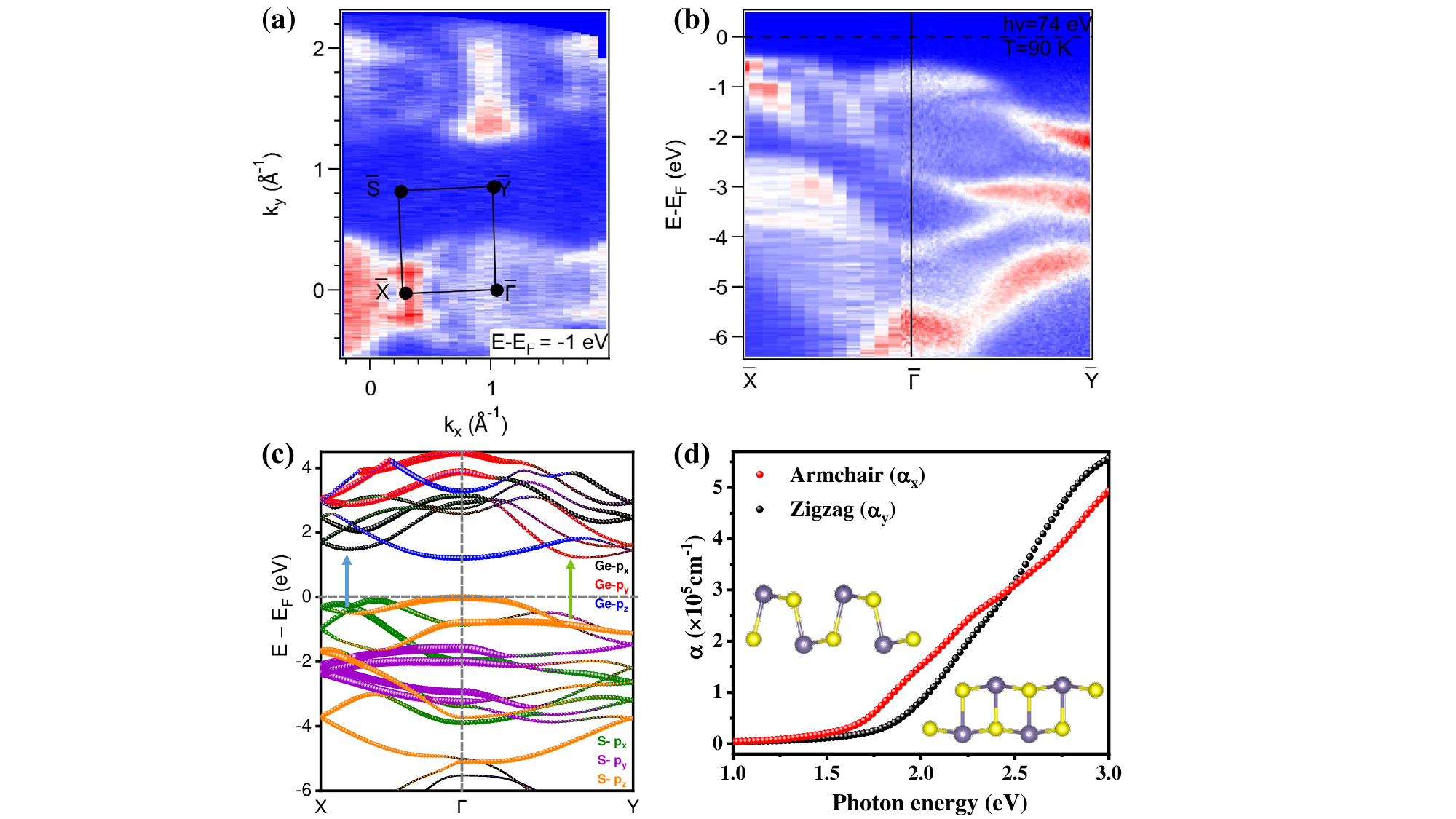}}
\renewcommand{\thefigure}{S\arabic{figure}}
\caption{ARPES (photon energy 74 eV) and DFT: (a) Constant energy contour at a binding energy of $−$1 eV. Rectangular symmetry is consistent with the orthorhombic crystal structure. (b) ARPES spectra of $in \ situ$ cleaved GeS crystal at 93 K showing anisotropic band dispersion along $\bar{\Gamma}-\bar{X}$ and $\bar{\Gamma}-\bar{Y}$ direction. (c) Orbital projected band structure calculated using DFT framework depicting a band gap of 1.2 eV. (d) Calculated absorption coefficients of GeS as a function of photon energy along armchair and zigzag crystal orientation.
\label{arpes}}
\end{figure}

The optical-absorption spectrum of GeS was calculated using density functional theory to reveal the anisotropy in electronic structure, as shown in \fig{arpes}(d). The absorption coefficient $\alpha_x$ begins to increase from $\sim$1.6 eV indicating that the optical-absorption edge is around 1.6 eV which corresponds to the transition marked by blue arrow in \fig{arpes}(c). In contrast, the optical absorption edge along y direction ($\alpha_y$) is at $\sim$ 1.75 eV corresponding to the transition marked by green arrow in \fig{arpes}(c). Also, the absorption intensity is weak compared to $\alpha_x$ within the same energy regime. The absorption coefficient reaches a relatively larger value of 3.0 $\times$ 10$^5$ at around 2.5 eV. This large value of the absorption coefficient comes from the optically allowed transition above 1.6 eV. The calculated value of the optical-absorption edge is consistent with the previously reported experimental absorption spectroscopy measurements of bulk GeS \cite{2016-nanoresearch,materialsadvances-2020}.

\newpage
\begin{table}[h]
\renewcommand{\thetable}{S\Roman{table}}
\centering
\caption{Comparison of mobilities of different vdW materials on SiO$_2$ back-gate at 300K.}
\begin{tabular}{|c|c|c|c|c|c|}
\hline
\textbf{Material} & \textbf{Structure} & \textbf{In-Plane} & \textbf{Mobility} & \textbf{Mobility Ratio} & \textbf{Reference} \\
&  & \textbf{Anisotropy} & \textbf{(cm$^{2}$/V·s)} & \textbf{$\mu_{AC}/\mu_{ZZ}$} & \\ [1.5ex]
\hline
 MoS$_2$ & Hexagonal & No & $0.1$ - $10$ & $-$ & \cite{NatNano_2011_MoS2} \\ [1.5ex]
\hline
WS$_2$ & Hexagonal & No & $5$ -$ 50$ & $-$ & \cite{ACS-Nano_2014_WS2}  \\[1.5ex]
\hline
WSe$_2$ & Hexagonal & No & $\sim$ 250 & $-$ & \cite{NanoLett_2012_WSe2}  \\[1.5ex]
\hline
BP & Orthorhombic & Yes & $\sim$ 286 & $\sim$ 2 & \cite{ACSNano_2014_BP},\cite{NatCommun_2014_BPreview} \\[1.5ex]
\hline
 SnS& Orthorhombic & Yes & $\sim$ 20 & $\sim$ 0.6 & \cite{ACSNano_2017_SnS} \\[1.5ex]
\hline
SnSe & Orthorhombic & Yes & $\sim$ 1.5 & $\sim$ 3  & \cite{PRB_2015_SnSe_SnS} \cite{NanoRes_2015_SnSe} \\[1.5ex]
\hline
 ReS$_2$& Triclinic& Yes & 0.1 - 15 & $\sim$ 3.1 & \cite{NatCommun_2015_ReS2} \\[1.5ex]
\hline
 TiS$_3$& Monoclinic &Yes & $\sim$ 80 & $\sim$ 8 & \cite{AdvMater_2015_TiS3} \\[1.5ex]
\hline
 GeS& Orthorhombic &Yes& 0.038 & $\sim$ 3.4 & This Work \\[1.5ex]
\hline
\end{tabular}
\end{table}

\begin{table}[h]
\renewcommand{\thetable}{S\Roman{table}}
\centering
\caption{Summary of transport measurements for GeS devices measured along the Armchair (AC) and Zigzag (ZZ) directions.}
\begin{tabular}{|c | c | c | c | c | c|}
\hline
\textbf{Device} & \textbf{AC} & \textbf{ZZ} & \textbf{$\mu_{AC}$} & \textbf{$\mu_{ZZ}$} & \textbf{$\mu_{AC}/\mu_{ZZ}$} \\
 & \textbf{On-Off Ratio} & \textbf{On-Off Ratio} & \textbf{(cm$^2$V$^{-1}$s$^{-1}$)} & \textbf{ (cm$^2$V$^{-1}$s$^{-1}$)} &  \\
\hline
1 & $5 \times 10^{2}$ & $1.5 \times 10^{2}$ & 0.038 & 0.011 & 3.45 \\[1.5ex]
2 & $9 \times 10^{2}$ & $3.1 \times 10^{2}$ & 0.043 & 0.013 & 3.30 \\[1.5ex]
3 & $2.5 \times 10^{2}$ & $1\times 10^{2}$ & 0.022 & 0.008 & 2.75 \\[1.5ex]
4 & $4.5 \times 10^{2}$ & $1.2 \times 10^{2}$ & 0.033 & 0.09 & 3.7 \\[1.5ex]
\hline
\end{tabular}
\label{tab:GeS_devices}
\end{table}

\begin{table}[h]
\renewcommand{\thetable}{S\Roman{table}}
\centering
\small
\setlength{\tabcolsep}{3pt}
\caption{GeS/MoS$_2$ heterostructure diodes: directional rectification and antiambipolarity.}
\setlength{\tabcolsep}{3pt}
\begin{tabular}{|c | cc | c | c | cc|}
\hline
{\textbf{Device}} &
\multicolumn{2}{c|}{\textbf{Rectification ratio}} &{\textbf{AC/ZZ}} &
\textbf{Antiambipolar} &
\multicolumn{2}{c|}{\textbf{Antiambipolar}} \\
 & \multicolumn{2}{c|}{} & & \textbf{ Peak Voltage [V]} &
\multicolumn{2}{c|}{\textbf{ Peak Current [nA]}} \\
\hline
 & \textbf{AC} & \textbf{ZZ} &  &  & \textbf{AC} & \textbf{ZZ} \\
\hline
1 & 41 & 19 & 2.2 & $-$4.5& 2.23 & 0.55 \\[1.5ex]
2 & 53 & 28 & 1.89 &   $-$12 & 3.84 & 1.15 \\[1.5ex]
3 & 37 & 15 & 2.47 &  1.5  & 1.57 & 0.43 \\[1.5ex]
\hline
\end{tabular}
\label{tab:hetero_diode_summary}
\end{table}

\begin{figure}[ht]
\centerline{\includegraphics[scale=0.55, clip]{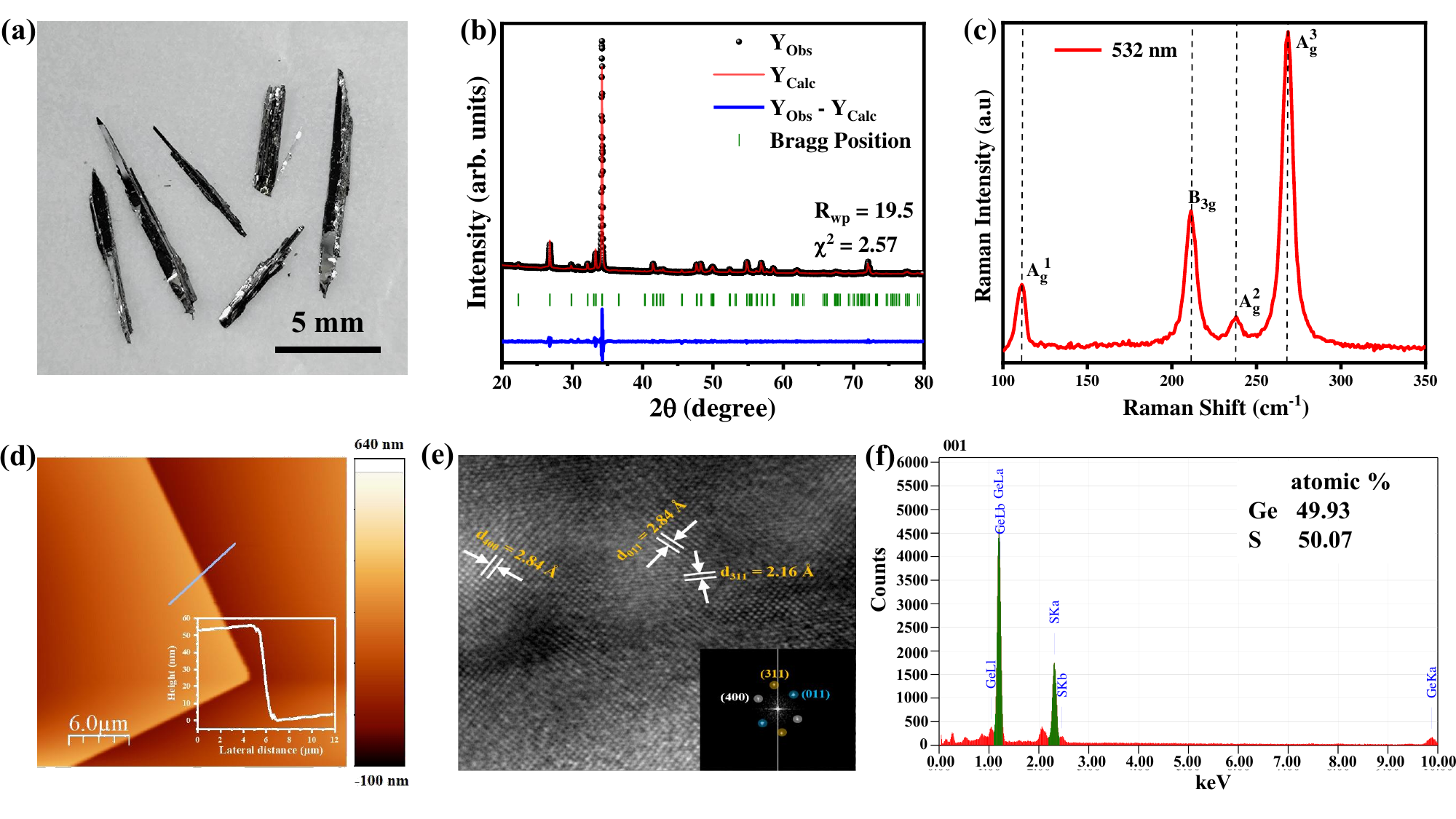}}
\renewcommand{\thefigure}{S\arabic{figure}}
\caption{(a) Optical image of the as-grown GeS single crystals. (b) Rietveld refined powder X-ray diffraction pattern confirming the orthorhombic phase of GeS. (c) Unpolarized Raman spectrum of exfoliated GeS flake using 532 nm Laser source. (d) Atomic force microscope (AFM) imaging and the corresponding height profile across the line-cut of a multilayered GeS flake. (d) Transmission electron microscope (TEM) imaging along with the selected area electron diffraction (SAED) pattern showing different Bragg planes. (f) EDS (energy dispersive study) spectrum for the elemental analysis in GeS single crystal. 
\label{char}}
\end{figure}

\begin{figure}[ht]
\centerline{\includegraphics[scale=0.6, clip]{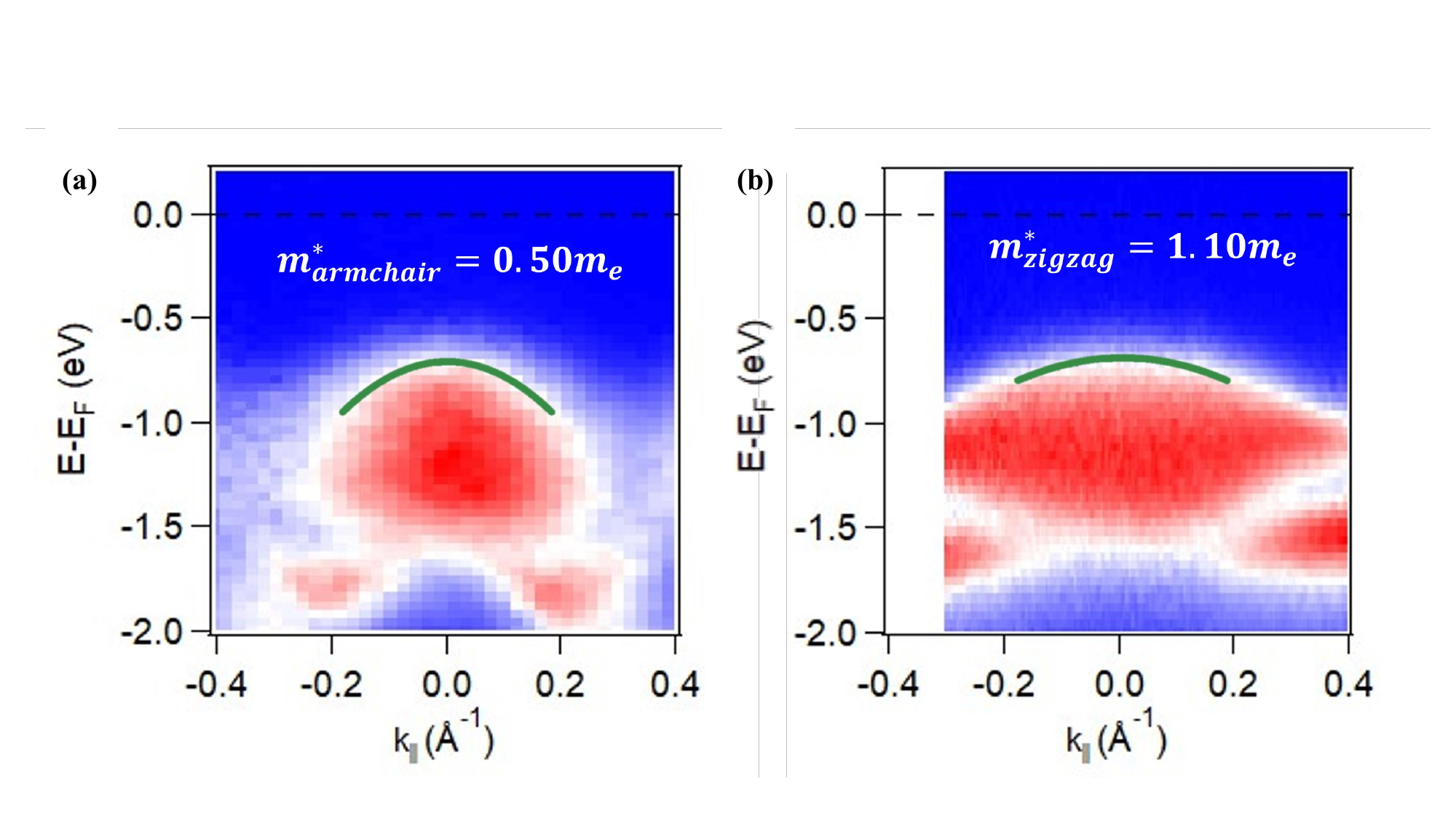}}
\renewcommand{\thefigure}{S\arabic{figure}}
\caption{High symmetric cuts along (a) armchair ($k_x$) and (b) zigzag ($k_y$) directions. The dotted lines in (a,b) depict the parabolic fit to the top-most valence bands. The effective masses of hole carriers along different orientations extracted from the fitting equations are shown in the figures.
\label{effmass}}
\end{figure}

\begin{figure}[ht]
\centerline{\includegraphics[scale=0.7, clip]{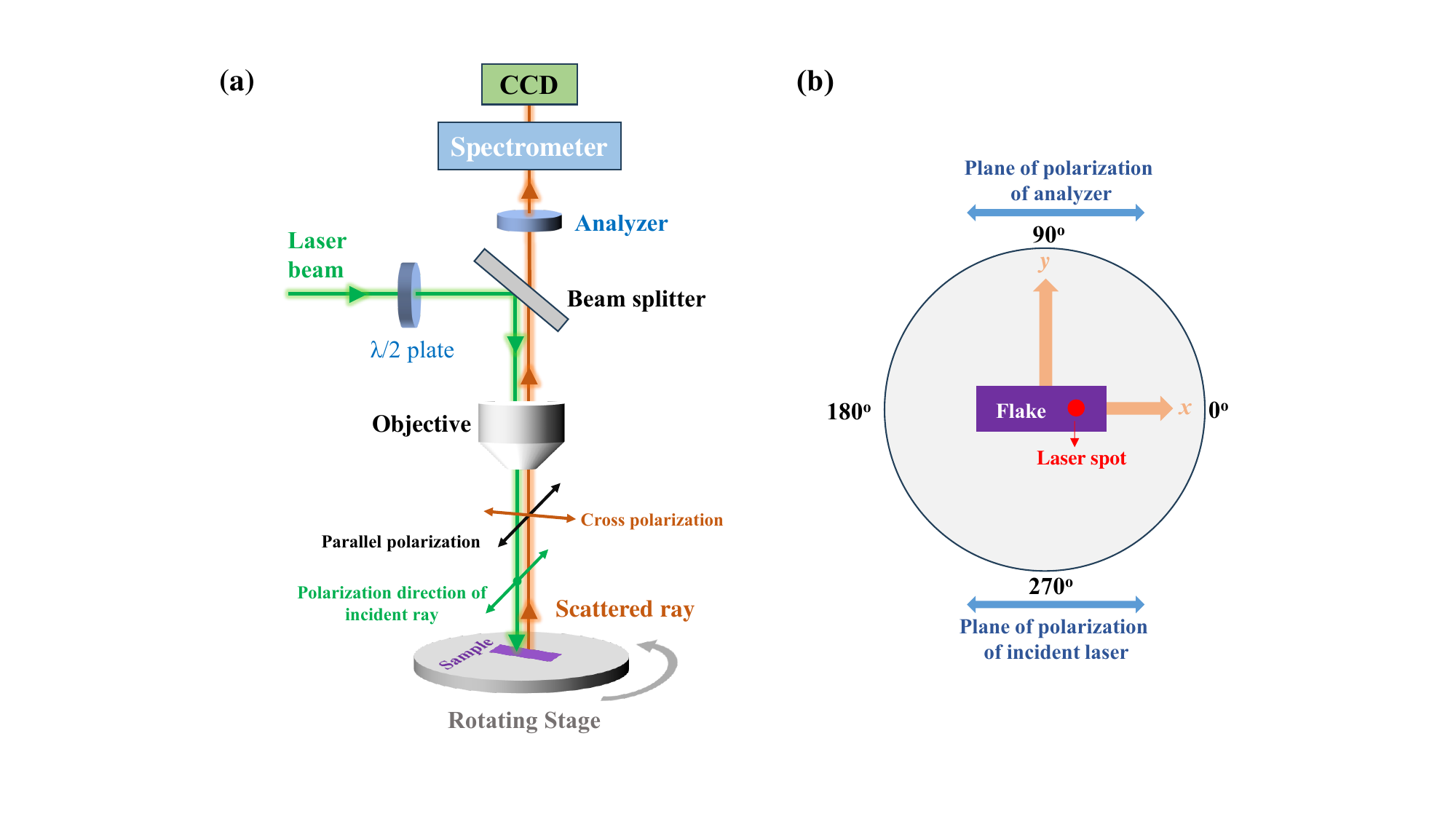}}
\renewcommand{\thefigure}{S\arabic{figure}}
\caption{Schematic of (a) the experimental setup of angle-resolved polarization Raman spectroscopy. (b) Top view of the parallel-polarization Raman setup in lab-frame.
\label{schematic}}
\end{figure}

\begin{figure}[ht]
\centerline{\includegraphics[scale=0.6, clip]{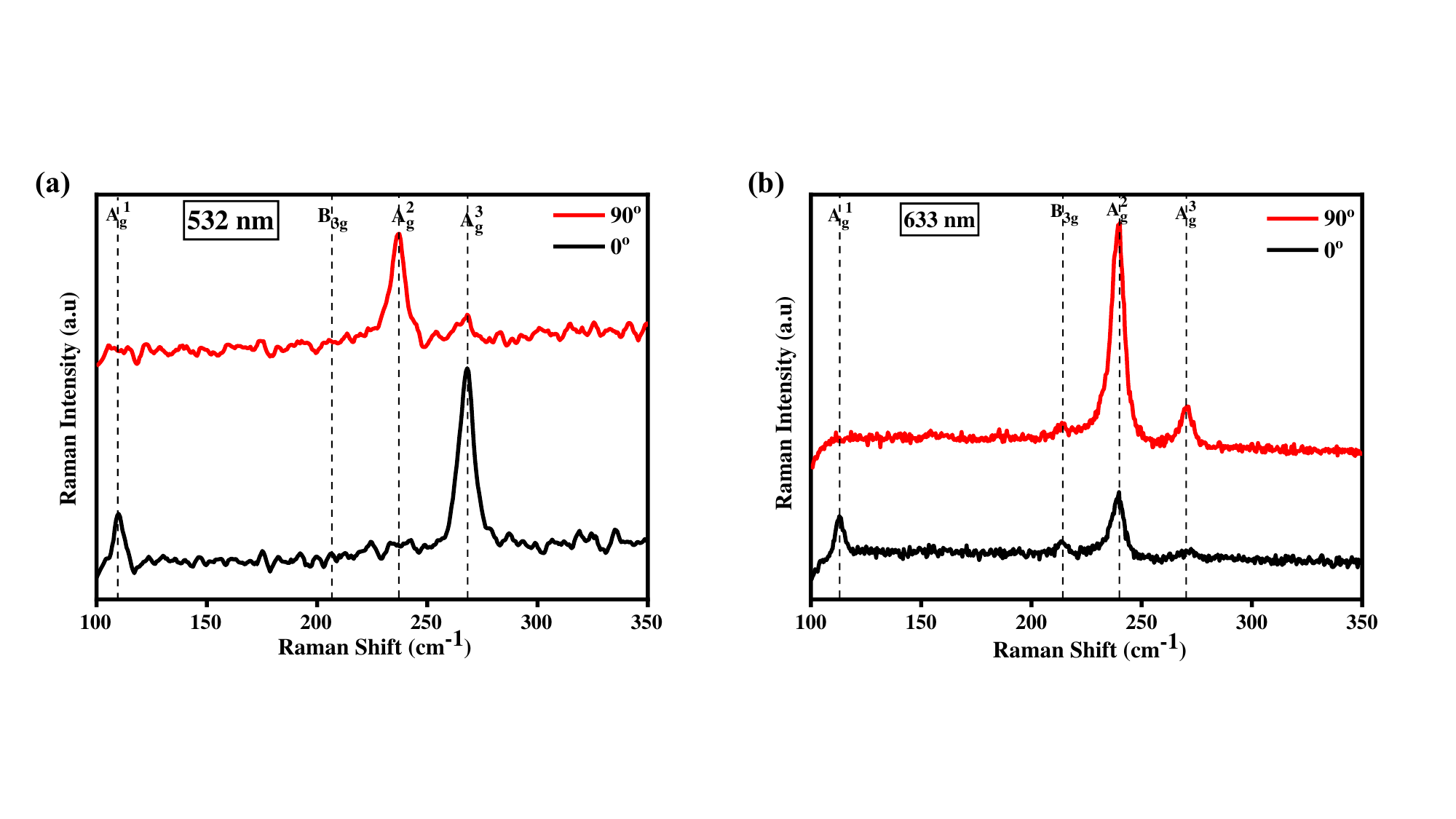}}
\renewcommand{\thefigure}{S\arabic{figure}}
\caption{(a) Polarized Raman spectra of GeS at 0\textdegree$ $ and 90\textdegree$ $ angle for (a) 532 nm, (b) 633 nm excitation under parallel polarization configuration. 
\label{Raman}}
\end{figure}

\begin{figure}[ht]
\centerline{\includegraphics[scale=0.6, clip]{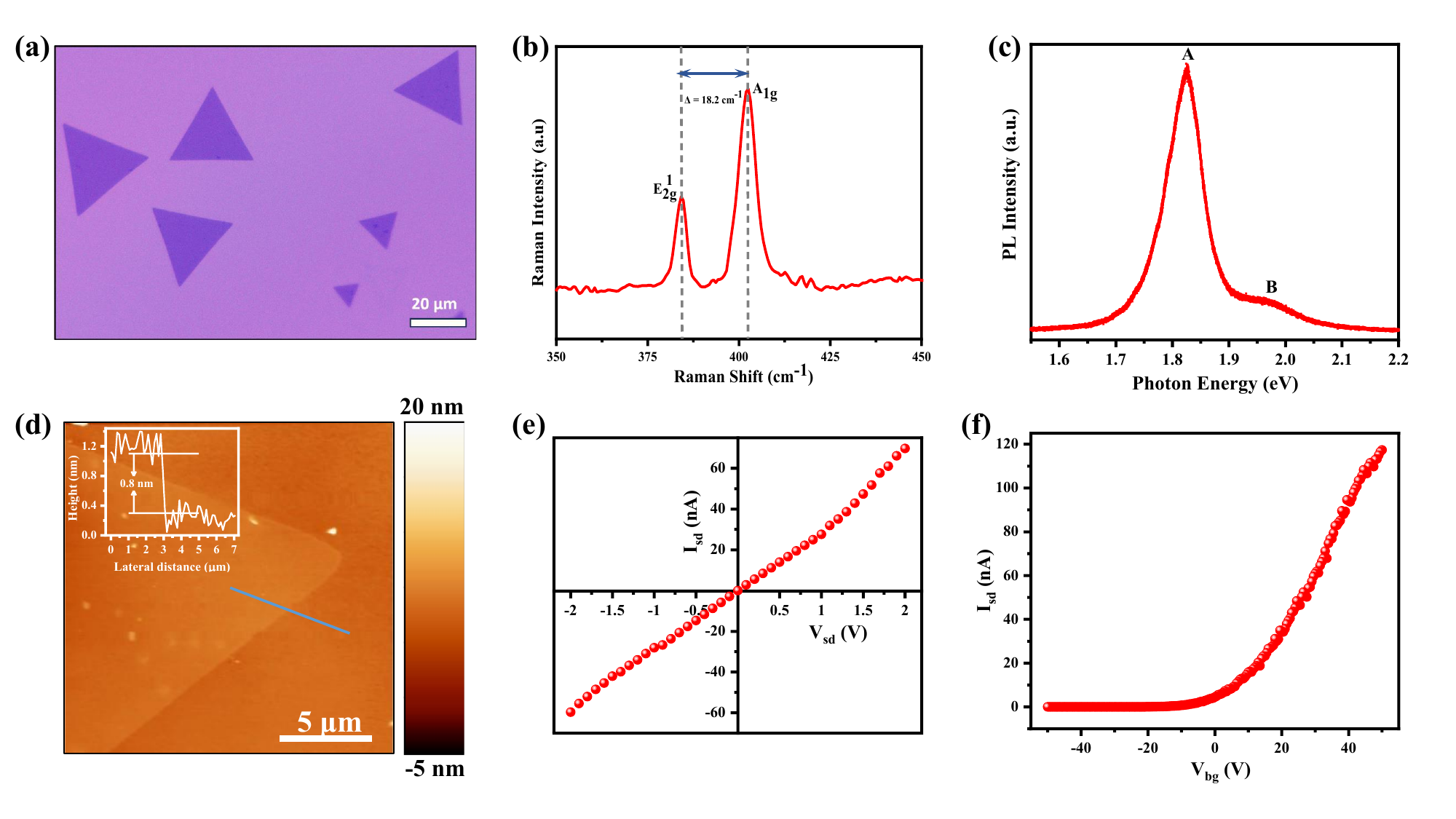}}
\renewcommand{\thefigure}{S\arabic{figure}}
\caption{(a) Optical image of CVD-grown monolayer MoS$_2$ flakes. (b) Raman spectrum of the as-grown flakes using 532 nm Laser source. The frequency difference between the two peaks confirms the monolayer thickness. (c) Room temperature PL spectra of monolayer MoS$_2$. (d) AFM image and the corresponding height profile of monolayer MoS$_2$ flake. (e) Linear $I_{\text{sd}}$-$V_{\text{sd}}$ characteristics. (d) $n$-type transfer characteristics of single-layer MoS$_2$ FET.     
\label{MoS2}}
\end{figure}

\begin{figure}[ht]
\centerline{\includegraphics[scale=0.8, clip]{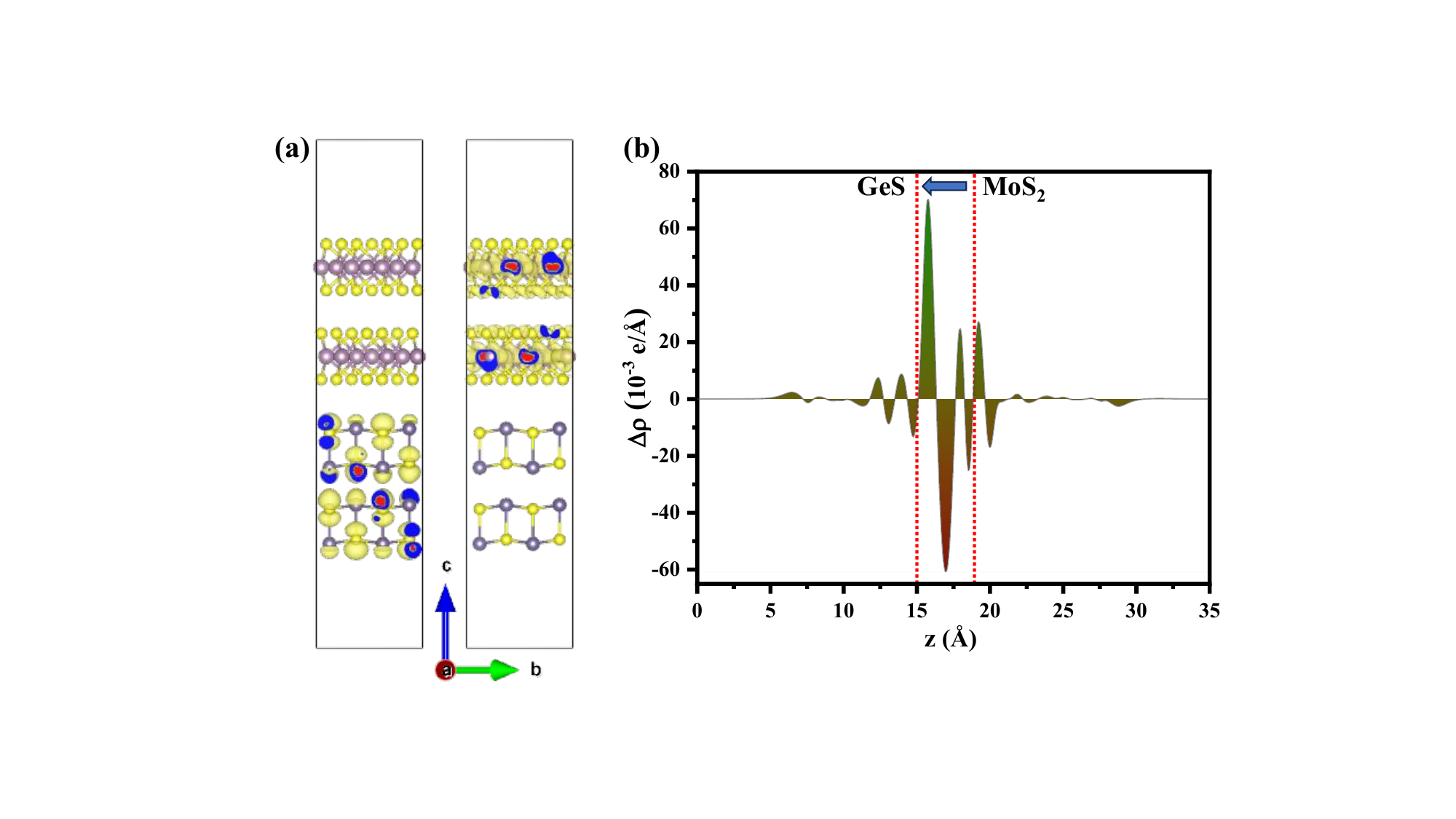}}
\renewcommand{\thefigure}{S\arabic{figure}}
\caption{(a) Partial charge density distributions at the (left) valence band maximum (VBM) and (right) conduction band minimum (CBM) for the GeS/MoS$_2$ heterostructure. The VBM states are primarily localized on the GeS layer, while the CBM states are concentrated on the MoS$_2$ layer, confirming a type-II band alignment that enables spatial separation of electrons and holes. (b) Planar-averaged charge density difference (CDD) along the out-of-plane direction at the interface of the GeS/MoS$_2$ vdW heterostructure. Positive and negative peaks indicate charge accumulation and charge depletion towards the GeS and MoS$_2$ layers, respectively. The central dotted region ($\sim$ 15–18 \AA) defines the GeS/MoS$_2$ interface, where the strongest charge redistribution occurs.    
\label{PARCHG}}
\end{figure}

\begin{figure}[ht]
\centerline{\includegraphics[scale=0.55, clip]{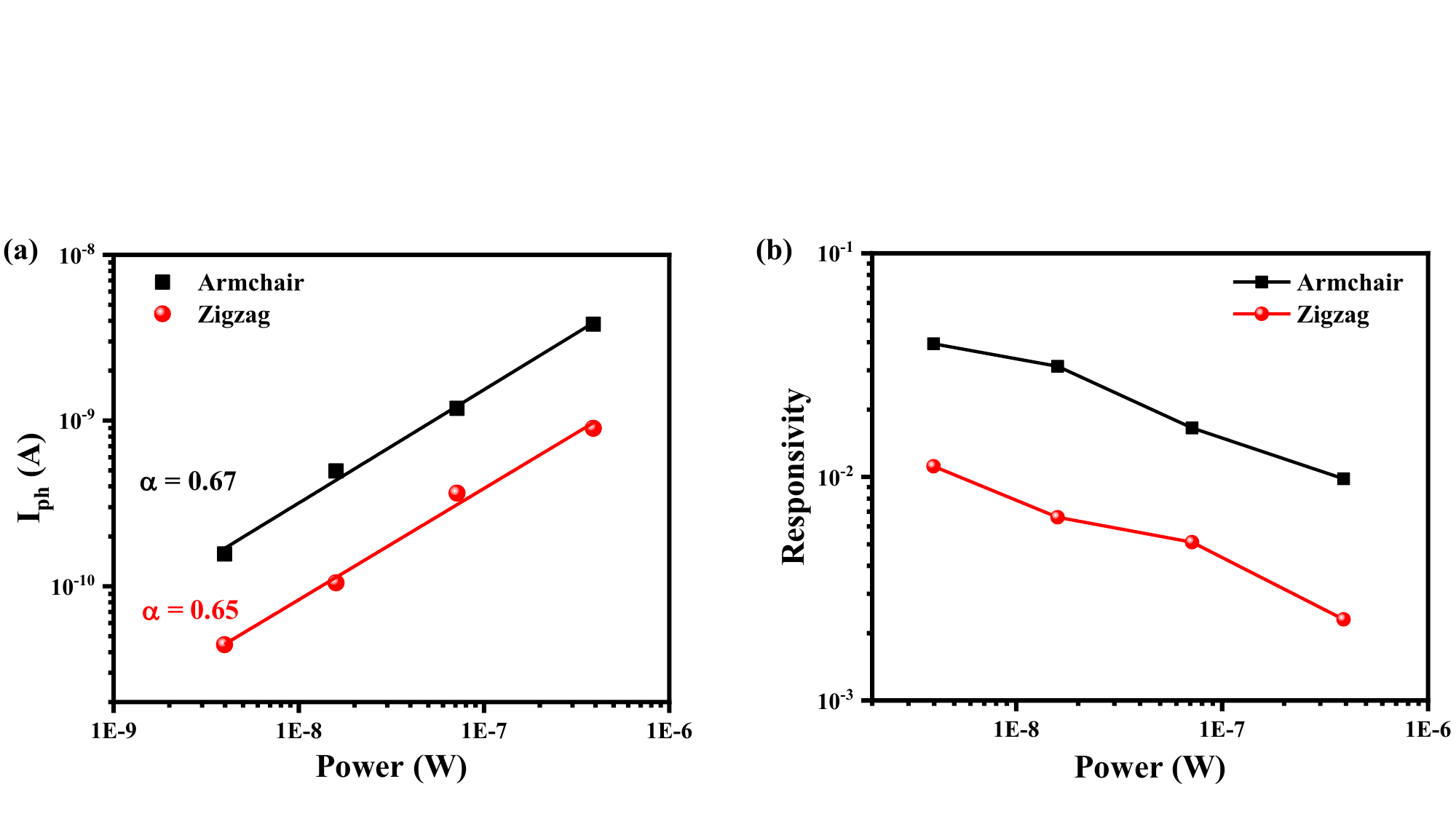}}
\renewcommand{\thefigure}{S\arabic{figure}}
\caption{(a) Photocurrent of GeS/MoS$_2$ vdWH as a function of Laser power along armchair and zigzag direction for 633 nm Laser source. (b) Responsivity $vs.$ Laser power calculated along armchair and zigzag orientations.     
\label{responsivity}}
\end{figure}